\documentclass[aps,prb,reprint,superscriptaddress]{revtex4-1}
\usepackage{graphicx}
\begin{document}


\title{Magnon dispersion in Ca$_2$RuO$_4$: impact of spin-orbit coupling and oxygen moments}


\author{S. Kunkem\"oller}
\affiliation{$I\hspace{-.1em}I$. Physikalisches Institut,
Universit\"at zu K\"oln, Z\"ulpicher Str. 77, D-50937 K\"oln,
Germany}

\author{E. Komleva}
\affiliation{Ural Federal University, Mira Street 19, 620002 Ekaterinburg, Russia}

\author{S. V. Streltsov}
\affiliation{M. N. Miheev Institute of Metal Physics of Ural Branch of Russian Academy of Sciences, 620137 Ekaterinburg, Russia}
\affiliation{Ural Federal University, Mira Street 19, 620002 Ekaterinburg, Russia}

\author{S. Hoffmann}
\affiliation{$I\hspace{-.1em}I$. Physikalisches Institut,
Universit\"at zu K\"oln, Z\"ulpicher Str. 77, D-50937 K\"oln,
Germany}

\author{D.~I.~Khomskii}
\affiliation{$I\hspace{-.1em}I$. Physikalisches Institut,
Universit\"at zu K\"oln, Z\"ulpicher Str. 77, D-50937 K\"oln,
Germany}

\author{P. Steffens}
\affiliation{Institut Laue Langevin, 6 Rue Jules Horowitz BP 156, F-38042 Grenoble CEDEX 9, France}

\author{Y. Sidis}
\affiliation{Laboratoire L\'eon Brillouin, C.E.A./C.N.R.S.,
F-91191 Gif-sur-Yvette CEDEX, France}

\author{K. Schmalzl}
\affiliation{J\"ulich Centre for Neutron Science JCNS, Forschungszentrum J\"ulich GmbH, Outstation at ILL, 38042 Grenoble, France}

\author{M. Braden}\email[e-mail: ]{braden@ph2.uni-koeln.de}
\affiliation{$I\hspace{-.1em}I$. Physikalisches Institut,
Universit\"at zu K\"oln, Z\"ulpicher Str. 77, D-50937 K\"oln,
Germany}



\date{\today}

\begin{abstract}

The magnon dispersion of Ca$_2$RuO$_4$ has been studied by polarized and unpolarized neutron scattering experiments on crystals containing 0, 1 and 10~\% of Ti. The entire dispersion of transverse magnons can be well
described by a conventional spin-wave model with interaction and anisotropy parameters that agree with density functional theory calculations.
Spin-orbit coupling strongly influences the magnetic excitations, which is most visible in large energies of the magnetic zone-center modes
arising from magnetic anisotropy. We find evidence for a low-lying additional mode that exhibits strongest scattering intensity near the
antiferromagnetic zone center. This extra signal can be explained by a sizable magnetic moment of 0.11 Bohr magnetons on the apical oxygens parallel to the Ru moment, which is found in the density functional theory calculations. The energy and the signal strength of the additional branch are well described by taking into account this oxygen moment with weak ferromagnetic coupling between Ru and O moments.

\end{abstract}

\pacs{}

\maketitle

\section{Introduction}
Ca$_2$RuO$_4$ (CRO) is the Mott insulating\cite{Nakatsuji1997} end member of the series Ca$_{2-x}$Sr$_x$RuO$_4$, which possess a rich diversity of structural, magnetic and transport properties\cite{Nakatsuji2000a,Friedt2001,Carlo2012,Nakatsuji2004}. Sr$_2$RuO$_4$, the other end member, is a spin-triplet superconductor with broken time-reversal symmetry \cite{Y.Maeno1994,Ishida1998,Luke1998,Maeno2012}. The metal-insulator transition in CRO at $T_{MIT}=360$~K \cite{Nakatsuji1997} goes along with severe structural distortions \cite{Alexander1999,Friedt2001}, the RuO$_6$ octahedrons flattens, which is also apparent in jumps of the lattice constants in the order of 0.1~\AA. These structural distortions are enhanced till the onset of antiferromagnetic order \cite {Braden1998}. The nowadays widely used picture assumes that an orbital ordering is associated with the structural changes \cite{Mizokawa2001,Gorelov2010,Sutter2016}. The 4$d_{xy}$ orbitals become doubly occupied and the  4$d_{xz,yz}$ singly occupied resulting in flattened octahedrons and a $S=1$ state. In the past the nature of the Mott transition of this multiband system with four $d$ electrons on the Ru site was intensively discussed \cite{Mizokawa2001,Jung2003,Hotta2001,Puchkov1998,Liebsch2007,Liebsch2003,Anisimov2002}. Recently, it was proposed, that spin-orbit coupling (SOC) in this 4$d$ system is strong enough to change the multiplet structure and couples $S$ and $L$ to $j$ resulting in a nonmagetic $j=0$ ground state. The occurring of a magnetic order was proposed to be of a singlet magnetism type (see e.g. Sec. 5.5. in Ref. \cite{Khomskii2014}), which was called excitonic magnetism in Ref. \cite{Khaliullin2013}, and a special type of magnon dispersion was predicted in this theory \cite{Akbari2014}. The main branches of the dispersion obtained by inelastic neutron scattering (INS) experiments, however, could be successfully described with a conventional Heisenberg model \cite{Kunkemoeller2015,Jain} and strongly disagree with $j=0$ calculations \cite{Akbari2014}. Nevertheless, neutron scattering experiments revealed features and additional magnetic scattering not describable with a simple square-lattice antiferromagnet. The main in-plane transverse modes exhibit a large anisotropy gap, underlining the impact of broken tetragonal symmetry in combination with SOC. Furthermore, an additional signal was detected in the neutron scattering experiments at low and at high energies \cite{Kunkemoeller2015,Jain}, that cannot be explained by the two transverse magnon branches.

Here we present a study of the magnon dispersion using INS with and without polarization analysis on crystals containing 0, 1, and 10~\% Ti combined with density functional theory (DFT) calculations. The full dispersion of magnetic excitations in CRO reveals a very strong impact of SOC, and oxygen moments that lead to an additional branch.

\section{Experimental}
  \begin{figure}
  \includegraphics[width=0.9\columnwidth]{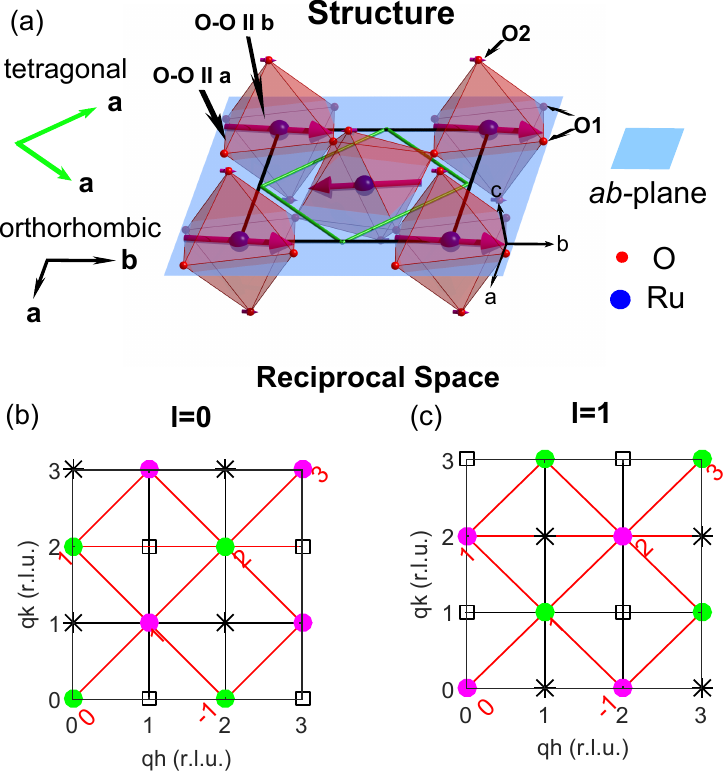}%
  \caption{\label{explanations}(color online) Crystal and magnetic structure of CRO and the associated reciprocal space. In (a) one layer of the crystal and magnetic structure is shown. The Ru (blue balls) sit in octahedrons of oxygen (red balls).  The ab-plane is indicated in light blue. In black and green the orthorhombic and tetragonal cells are shown. The pink arrows denote the magnetic moments on the Ru and O sites. At the tip of the low right Ru moment, the directions of the different polarizations of the magnon modes are indicated by black arrows and some labellings of atoms and atom distances are given. In (b,c) the (hkl) planes in reciprocal space are illustrated for l=0 and l=1, respectively. The orthorhombic and tetragonal cells are drawn in black and red, respectively. Green circles denote tetragonal zone centers and pink circles zone boundaries. Black squares (stars) mark antiferromagnetic zone centers of the $A$- ($B$)-centered antiferromagnetic order.}
  \end{figure}

Crystal growth of insulating Ca$_{2-x}$Sr$_x$RuO$_4$ is severely hampered by the metal-insulator transition occurring in pure CRO slightly  above room temperature, $T_{MIT}=360$~K. The space group does not change at this first-order phase transition, but there are sizable jumps in the lattice-parameters, in particular for $c$, with $\Delta c\sim 0.2$~\AA \cite{Friedt2001}. Therefore, upon cooling the crystals after the growth process, they tend to crack and only small pieces of mm$^3$ size can be recovered. Because inelastic neutron scattering experiments require samples of $\sim$cm$^3$ size, we substituted a small amount of Ru by Ti in order to broaden the phase transition and to be able to cool the crystals through the metal-insulator transition without breaking them into small pieces. Crystals with a mass of 0.5~g to 1~g  were obtained with only 1~\% Ti substitution. We were also able to obtain two larger pieces of pure CRO, one with a mass of 0.45~g and one with 0.3~g, but these crystals posses a bad mosaic spread of 5$^\circ$, which most likely is the reason why they were not destroyed upon cooling. Nevertheless, these crystals are suitable for some inelastic neutron scattering studies.

The Ca$_2$Ru$_x$Ti$_{1-x}$O$_4$ crystals with $x=0, 1, and 10~$\% used in this neutron scattering study (labeled 0Ti, 1Ti, 10Ti, respectively) were grown by the floating-zone method in a Canon Machinery Inc. SC1-MDH11020-CE furnace equipped with two 2000~W halogen lamps and a coldtrap following the procedure described in references \cite{Fukazawa2000,Kunkemoeller2016}. CaCO$_3$, RuO$_2$ and TiO$_2$ were mixed in stoichiometric ratios and a Ru excess of 32~\% was added. The powder was mixed and reacted for 24~h with an intermediate grinding. Then a rod was pressed and sintered at 1350~$^\circ$C. A growth speed of 17~mm/h and a feed-rod speed of 20~mm/h were used and the atmosphere contained 90~\% Ar and 10~\% O at a total pressure of 9~bar. Phase purity was checked by x-ray powder diffraction from crushed parts of the single crystals, which indicates no impurity phases. The lattice constants at room temperature were obtained by LeBail fits of these powder pattern using the FullProf Suite\cite{Rodriguez-Carvajal1993}. Single-crystal x-ray diffraction experiments were performed on a Bruker X8 Apex diffractometer equipped with a charge-coupled-device detector. Details of this structure determination and a detailed list of the refined parameters can be found in the Appendix A. Magnetization measurements were performed with a superconducting quantum interference device (SQUID) magnetometer from Quantum Design, and the resistivity was measured with a standard four-point method. Spin wave calculations were performed using SpinW \cite{toth2015}.

Elastic and inelastic neutron scattering experiments were performed with the cold triple-axis spectrometer (TAS) 4F1 at the LLB, with the thermal TAS IN3 at the ILL and with the polarized thermal TAS IN22 at the ILL. The polarized neutron scattering experiments were performed using Heusler (111) monochromator and analyzer crystals. A PG filter was inserted on the scattered beam before the neutron-spin flipper and the monitor was put on the incoming beam between monochromator and sample. A set of Helmholtz coils was used to produce the guide field and the sample was zero field cooled (less than 2~G) in an orange-type cryostat. A mounting of 3 crystals with a total mass of 2.5~g containing 1~\% Ti (1Ti) was used for the polarized neutron scattering experiment. The sample was twinned with non-equal twinning-fractions of 2.5:1 determined by scanning the orthorhombic (200) and (020) reflections. The $a,b$ plane [see Fig.~\ref{explanations}(a)] was chosen as the scattering plane in order to efficiently integrate the inelastic signal along the vertical direction, where the resolution is poor. We used the standard coordinate system in polarized neutron scattering \cite{Chatterji2005}. \textbf x is parallel to the scattering vector (\textbf Q), \textbf y and \textbf z are perpendicular to \textbf Q. While \textbf y is in the scattering plane, \textbf z is perpendicular to it. Therefore, \textbf y lies in the $a,b$~plane of the crystal and \textbf z parallel to the $c$~axis.  In neutron scattering only magnetic moments perpendicular to \textbf Q intervene. Therefore, the magnetic scattering intensities sense a geometry factor sin$^2$($\alpha$) with $\alpha$ being the angle between \textbf Q and the magnetic moment, which corresponds to either the static ordering moment in diffraction or to the oscillating moment in a magnon. With longitudinal neutron polarization  six different channels can be analyzed: three spin-flip (SF) and three non-spin-flip (nSF) channels. While phonons always contribute to the nSF channel there is an additional selection rule for magnetic scattering. The magnetic component parallel to the direction of the neutron polarization contributes to the nSF channel, while the components perpendicular to the neutron polarization generate SF scattering. By combining this polarization rule with the geometry factor one can distinguish the
different magnetic components. In the geometry we use for the polarized neutron scattering experiment we see the $c$ polarized modes in the SFy channel and they do not loose intensity due to the geometry factor, because the scattering vector is always parallel to the $a,b$ plane and thus perpendicular to $c$.   In contrast the SFz channel contains the in-plane modes, the transverse and longitudinal ones, but weighted with the geometry factor.   For example at $\textbf Q=(2,1,0)$ the geometry factor for the transverse mode is sin$^2(\alpha)=0.2$ and that for the longitudinal mode sin$^2(\alpha)=0.8$. At $\textbf Q=(1,2,0)$ this ratio is inverted [see Fig. \ref{explanations}(b,c)]. Respecting the twinning ratio of 2.5:1, we expect that the intensities at $\textbf Q=(2,1,0)$ and $\textbf Q=(1,2,0)$ have a ratio of 0.6 for the transverse mode. For the longitudinal mode this ratio is inverted. Thus, it is possible to distinguish between transverse modes and longitudinal modes by comparing scattering at properly chosen \textbf Q. Because the polarization of the neutron beam is not perfect, one has to correct the intensities for the finite flipping-ratio (FR) \cite{Chatterji2005,Qureshi2012}. Thereby, the magnetic signals are obtained from the intensities of the different SF channels corrected for the FR:
\begin{equation}
  I(M_{y,z})=\frac{FR+1}{FR-1}[I(SF_x)-I(SF_{y,z})]
 \end{equation}
 The FR of our experiment on IN22 amounts to 12, which is determined by comparing the signals of rocking scans on the (200) Bragg-reflection in the SF and nSF channels.

 On the cold TAS 4F1 crystals with different Ti content were used, 0Ti and 10Ti. 0Ti has a mass of 0.45~g and a twinning ratio of 1:1 and 10Ti a mass of 0.93~g and a twinning ratio of 9:1. The scattering plane for both samples was chosen to be the [010]/[001]-plane. So the scattering plane for the other twin domains was the [100]/[001]-plane. For all 4F1 scans a pyrolytic graphite monochromator and analyzer were used, a cooled Be-filter was put on \textbf k$_f$ to suppress higher order contaminations. For the cooling of the samples an orange-type cryostat was used and $k_f$ was set to 1.55~\AA$^{-1}$ for all scans.

 With the IN3 TAS we analyzed the magnetic order of the crystal containing 1\% Ti used in the previous neutron scattering study\cite{Kunkemoeller2015} (1TiB). This crystal was found to exhibit a majority twin domain of 95\% and the experiment was performed in the [010]/[001] orientation used in the inelastic neutron scattering experiment.

\section{Results and discussion}
\subsection{Crystal structure of Ca$_2$Ru$_x$Ti$_{1-x}$O$_4$}

  \begin{figure}
  \includegraphics[width=0.9\columnwidth]{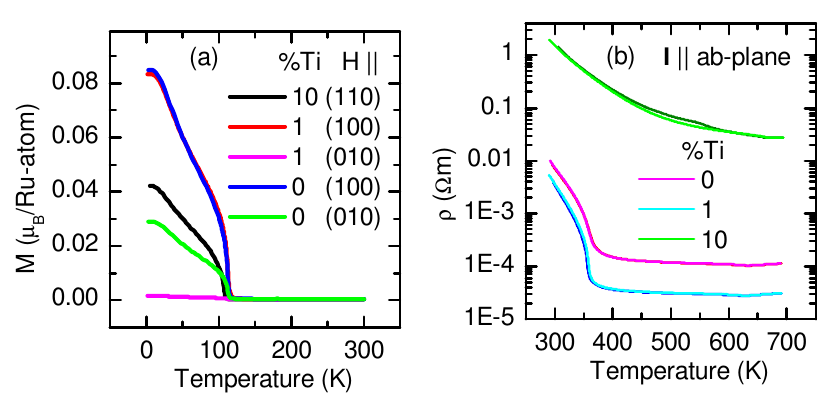}%
    \caption{\label{Magnetization}(color online) Temperature dependence of physical properties of CRO with different Ti content. (a) field-cooled magnetization curves  in 0.1~T measured on heating. (b) resistivity on heating in dark color and on cooling in bright color. Only the 10~\% Ti curves do not coincide.}
  \end{figure}

The crystal structure of CRO \cite{Braden1998} is similar to the structure of Sr$_2$RuO$_4$ of K$_2$NiF$_4$ type \cite{Randall1959}, where the Ru atoms sit in oxygen-octahedron cages, which are corner shared in the $a,b$~plane. In  CRO the octahedrons are rotated and tilted and they also become severely distorted. As a consequence, the tetragonal symmetry of Sr$_2$RuO$_4$ is reduced to the orthorhombic space group Pbca\cite{Braden1998}. One layer of the crystal structure is drawn in Fig.~\ref{explanations}, which also depicts the magnetic order with antiferromagnetic moments on Ru pointing along $b$. The orthorhombic unit cell is rotated by $45^\circ$ with respect to the tetragonal one and enhanced to: $\textbf{a}_{orth}=\textbf{a}_{tet}+\textbf{b}_{tet}$, $\textbf{b}_{orth}=\textbf{a}_{tet}-\textbf{b}_{tet}$, ${a}_{orth}\sim{b}_{orth}\sim\sqrt{2}{a}_{tet}$. Unless otherwise specified all notations refer to the orthorhombic lattice of the
 majority twin orientation. Our crystals usually posses two twin orientations, which are obtained by interchanging the ${a}_{orth}$ and ${b}_{orth}$ directions.

All crystals were examined by x-ray diffraction, magnetization and resistivity measurements in addition to the neutron scattering experiments described below.
Fig.~\ref{Magnetization} shows the magnetization and electric resistance data of the three Ti concentrations.
The temperature of the metal-insulator transition is 4~K lower for the sample containing 1~\% Ti than for the pure compound. The sample containing 10~\% Ti does not show indications for a phase transition upon cooling down to 80~K where the experimental limit of the increasing resistivity is reached. The absolute values of the resistivity curves have a large uncertainty because of the first-order structural phase transition. There the crystals tend to crack, which prohibits the current to flow through the hole sample. This effect has been frequently observed during a single measurement cycle. The resistivity is enhanced by a multiplication factor after passing the structural transition. As the samples have already passed this transition after the crystal growth, there are some cracks inside the sample, which is evident from measuring several pieces of the same crystal without obtaining the same room-temperature specific resistivity.

The magnetic and insulating properties in Ca$_{2-x}$Sr$_x$RuO$_4$ are closely related to the crystal structure. CRO is heavily distorted with respect to Sr$_2$RuO$_4$, which possesses the ideal structure of K$_2$NiF$_4$ type without structural distortions, but which already is close to such a structural instability\cite{Braden1998a,Braden2007,Braden1997}. In the layered ruthenates the distortions can be described as octahedron rotation around the $c$~direction and tilting around an in-plane axis. These distortions and the associated structural phase transitions result  from bond-length mismatch, and the distortions in CRO are induced by the chemical substitution of Sr by the isovalent, but smaller Ca, which is not able to fill the space between the oxygen octahedrons like Sr does. So the octahedrons start to rotate and tilt in order to reduce the coordination volume around Ca. By chemically substituting the smaller Ti for Ru, one expects a small decrease of these deformations, which is indeed realized. The  tilt and rotation angles of 10Ti are significantly smaller than the values determined in the samples containing only 0 and 1~\% Ti, see Appendix A, Tab.~\ref{tab:char}.

 \begin{figure}
  \includegraphics[width=0.79\columnwidth]{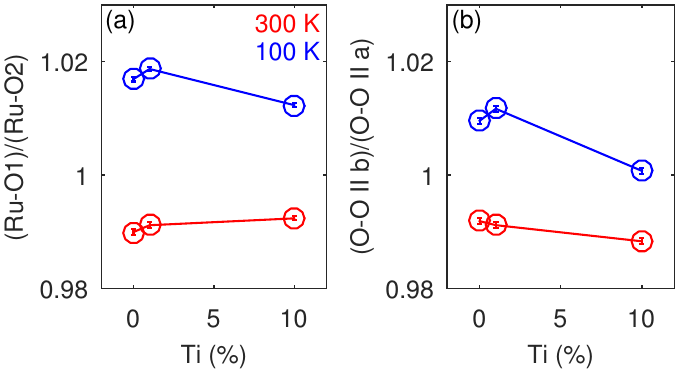}%
  \caption{\label{RuOOO} Temperature and Ti doping dependence of the RuO$_6$ octahedron distortion obtained by single crystal x-ray diffraction analyses; (a,b) presents the ratios of in-plane to out-of-plane RuO bond distances (flattening) and of the two octahedron edges parallel to $b$ and $a$, respectively. Both entities are enhanced in the insulating phase at low
    temperature due to the lifting of orbital degeneracy.}
  \end{figure}
Most interesting is the impact  of the Ti doping on the deformation of the RuO$_6$ octahedron, which is a fingerprint of the lifting of orbital degeneracy\cite{Braden1998,Friedt2001,Mizokawa2001}.
The structural change at the metal-insulator transition in pure CRO is characterized by a jump of the $c$~lattice-constant and by a flattening of the octahedron as it is visible in the Ru-O bond distance ratio\cite{Friedt2001}. Upon cooling into the insulating phase this octahedron flattening continues until it saturates at the antiferromagnetic phase transition. Slightly below room temperature the octahedron shape passes from elongated to flattened \cite{Friedt2001}. In all samples studied here, this crossover occurs below room temperature, thus at 100~K the octahedron shape is flattened for all Ti substitutions, see Fig.~\ref{RuOOO}. Due to flattening of the lattice and octahedron the $d_{xy}$ orbital shifts down in energy compared to the $d_{xz}$,$d_{yz}$  orbitals.
A similarly strong and anomalous temperature dependence is also observed for the ratio of the two O-O octahedron edge lengths parallel $a$ and $b$ which is related to the orthorhombic splitting
[$\epsilon=(b-a)/(a+b)$] \cite{Friedt2001}. At room temperature the octahedron is longer along $a$, while it is elongated along $b$ at low temperature. All these
effects can be attributed to a temperature dependent orbital ordering \cite{Braden1998,Friedt2001,Gorelov2010,Kunkemoeller2015}. The considerable elongation along $b$ at low temperature agrees with SOC and the alignment of the magnetic moment mainly along the $b$ direction \cite{Kunkemoeller2015}. Figure \ref{RuOOO} illustrates that these two distortions of the RuO$_6$ octahedron become suppressed by the 10~\% Ti substitution. This underlines the orbital ordering character of these distortions, which is obviously suppressed by Ti$^{4+}$ with
an empty $3d$ shell.

While the $a,b$~plane (or the average in-plane parameter) increases with high Ti content, the $c$~lattice-constant decreases at room temperature. The reduction of the $c$~lattice-constant, in hand with an increase of the $a,b$~plane, is the structural signature of the insulating state of CRO.
With 1~\% of Ti substitution there is only a small decrease in $T_{MIT}$ (Fig.~\ref{Magnetization}), with 10~\% of Ti substitution there are drastic effects. In the resistivity of 10Ti there are no indications for a sharp metal-insulator transition on cooling from 700 down to 80~K, where the upper experimental limit of the resistance experiment is reached. Note that the absolute values of the resistivity curves are prone to a large uncertainty because of the first order structural phase transition which causes cracks in the crystals. But the orders of magnitude larger resistivity of the 10Ti sample and the absence of the metal-insulator transition is unambiguous. This remarkable stabilization of the insulating state by only small amounts of Ti substitution is also seen in Ca$_3$Ru$_2$O$_7$ \cite{Ke2011}. Since the ionic radii of Ti and Ru are very similar, this suppression of the metallic state
seems to originate from the very effective suppression of the hopping. Also in Sr$_2$RuO$_4$ Ti substitution has a strong impact: it stabilizes spin-density wave
ordering associated with the Fermi-surface nesting of the pure material for only 2.5\%~Ti \cite{Braden2002}. Also in Sr$_3$Ru$_2$O$_7$ Ti substitution stabilizes a spin-density wave magnetic instability \cite{Steffens2009}.

\subsection{Magnetic structure of Ca$_2$Ru$_x$Ti$_{1-x}$O$_4$}

  \begin{figure}
  \includegraphics[width=0.85\columnwidth]{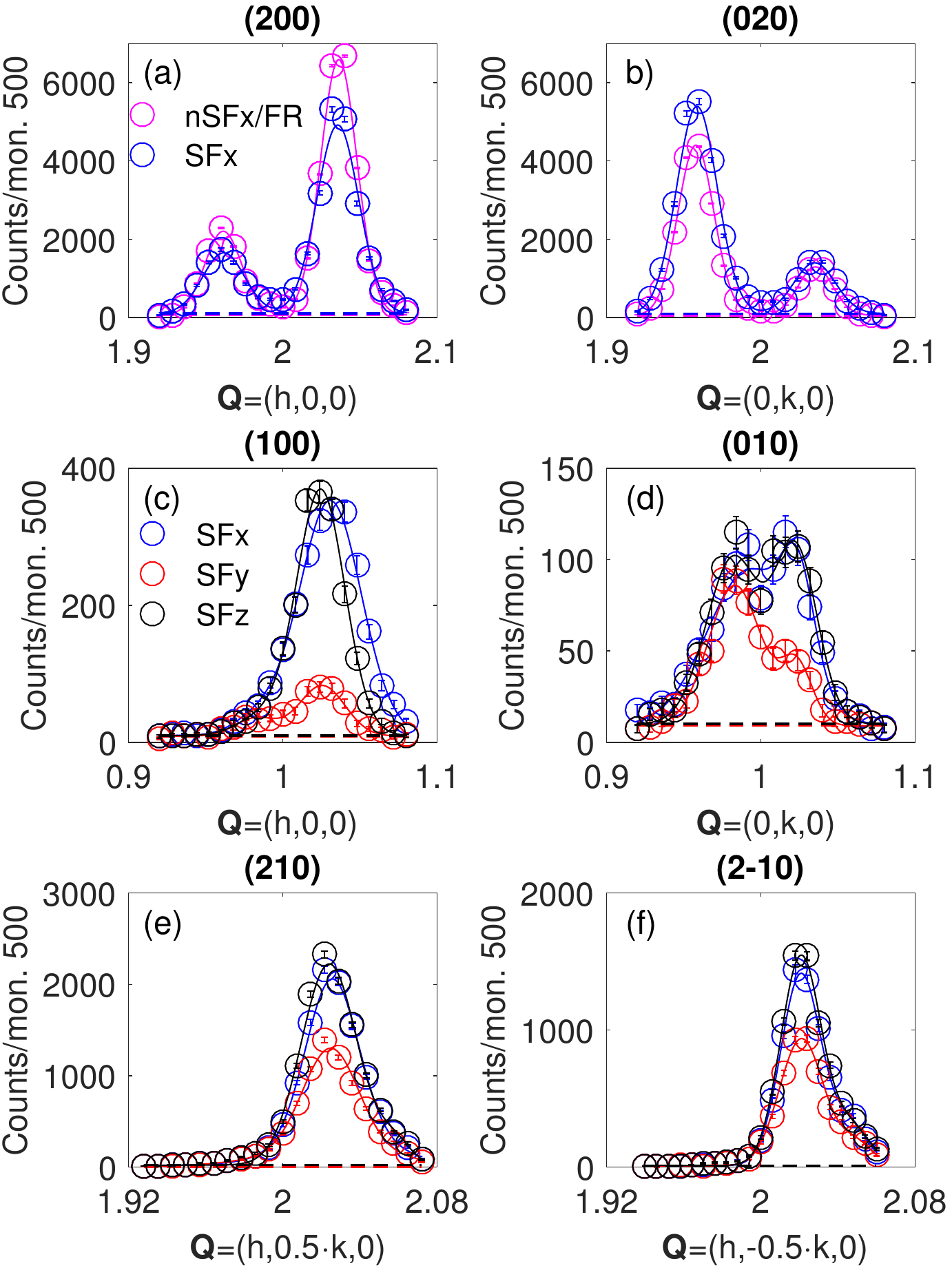}%
  \caption{\label{polel}(color online) Elastic neutron scattering scans with polarization analysis. The orthorhombic splitting is clearly visible in this elastic scans. (a,b) present the values of the nSFx (SFx) channel in magenta (blue) of longitudinal scans across the strong nuclear reflections (200) and (020), respectively. The values of the nSFx channel are divided by the FR. These elastic longitudinal scans show that both twins are well separated. (c-f) present data of the SFx (blue), SFy (red) and SFz (black) channels of scans across the magnetic (c) (100), (d) (010), (e) (210), and (f) (2-10) reflections. }
  \end{figure}

The magnetic structure of 1Ti was studied on IN22 with polarization analysis. The analysis of the magnetic structure of 1TiB on IN3 is described in Appendix B. Scans along the (200) and (020) reflections (Fig.~\ref{polel}) reveal that crystal 1Ti exhibits both twin-domain orientations in the ratio 2.5:1. Fig.~\ref{polel}(c-f) show the SF intensities in the three channels for scans across the (100), (010), (120) and (-120) magnetic reflections for crystal 1Ti. The magnetic structure of CRO has been previously determined by neutron powder diffraction \cite{Braden1998}. As illustrated in Fig.~\ref{explanations}(a) the magnetic moments are essentially aligned along the $b$~direction (corresponding to the tilt axis) with antiferromagnetic alignment between nearest neighbors. Due to the low symmetry of the crystal structure (Pbca) the Dzyaloshinski-Moriya interaction is finite and results in a canting of the moments along the $a$ direction. The canting of the moments along $a$ is parallel for all Ru sites in a single layer resulting in a net ferromagnetic moment
per layer. There are two different magnetic structures reported for CRO \cite{Friedt2001,Braden1998,Steffens2005}, which differ by the stacking of the single layer arrangement shown in Fig.~\ref{explanations}(a).
The magnetic structure of the main antiferromagnetic $b$ component is either $A$ or $B$-centered.  In the $A$-centered phase the two magnetic moments at (0,0,0) and (0,0.5,0.5) are parallel, in the $B$-centered one the two moments at (0,0,0) and (0.5,0,0.5) are parallel. The magnetic space groups are $Pbca$ ($A$-centered) and $Pbc'a'$ ($B$-centered). While in $Pbca$ the net ferromagnetic canted moments per layer cancel due to an antiferromagnetic stacking, the $B$-centered $Pbc'a'$ structure results in a total ferromagnetic moment, that can be measured with a magnetometer. There is also canting along $c$ direction in both magnetic structures.  This canting corresponds to an antiferromagnetic $c$ component that is $B$-centered in $Pbca$ and $A$-centered in $Pbc'a'$. This moment should, however, be small because the
main part of the ordered moment points along the tilt axis, and only microscopic methods can detect such an antiferromagnetic $c$ component.

The analysis of the magnetic Bragg peaks with polarization analysis first confirms that moments point along the $b$ direction and show a dominating
$B$-centered scheme for 1Ti and an almost exclusive $B$-centered scheme for 1TiB, see Appendix B. In contrast, small pure CRO samples
 show only the
$A$-centered scheme \cite{Steffens2005}.

In Fig.~\ref{Magnetization}(a) the magnetization on heating of samples with different Ti content is presented. The magnetic field of 0.1~T is applied in the $ab$ plane. The weak ferromagnetic component parallel to the  $a$ direction, resulting from the canting of the spins, dominates the magnetization below $T_N$. The fact that the magnetization of the lowest temperature is the highest, points to the $B$-centered phase, in which the ferromagnetic component does not cancel among two neighboring layers \cite{Braden1998}. Thus all samples contain a dominating $B$-centered phase in agreement with the neutron diffraction studies. The small crystal used for the magnetization measurement with 1~\% Ti content is essentially untwinned, while the sample containing no Ti is partially twinned with nonidentical twin fractions, in contrast to the much larger pure sample used for neutron scattering, for which the twinning ratio amounts to 1:1. If the field is applied along [1,1,0], the observable ferromagnetic component is reduced by a factor $1/{\sqrt{2}}$ but both twin domain orientations contribute (sample containing 10\% Ti). The highest total ferromagnetic component is observed in the pure sample, and the sample 1Ti already exhibits a reduced ferromagnetic component possibly stemming from a slightly reduced fraction of the $B$-centered phase.  We can conclude that CRO exhibits sizable moment canting resulting in an ordered magnetic component along the $a$ direction of $\sim0.08~\mu_B$ (canting angle $\alpha=3.5$~deg). The canting of the magnetic moment arises from a strong Dzyaloshinski-Moriya interaction that in turn stems from the strong SOC. Minimizing the static energy of a pair sensing only the Heisenberg interaction $J$ and the Dzyaloshinski-Moriya interaction, $-\textbf D\cdot\mathbf{S_i}\times\mathbf{S_{j}}-J\mathbf{S_i}\cdot\mathbf{S_{j}}$, yields the condition tan$(2\alpha)={D}/{J}$, and thus a rough estimate of the Dzyaloshinski-Moriya interaction in CRO: $D=0.06J$.

The magnetization curves further show a decrease of the N\'eel temperature with increasing Ti content, see Tab.~\ref{tab:char} in Appendix A, which can be explained by the dilution of the magnetic lattice by non-magnetic Ti.

\subsection{Polarization of magnon modes}

 \begin{figure}
 \includegraphics[width=0.88\columnwidth]{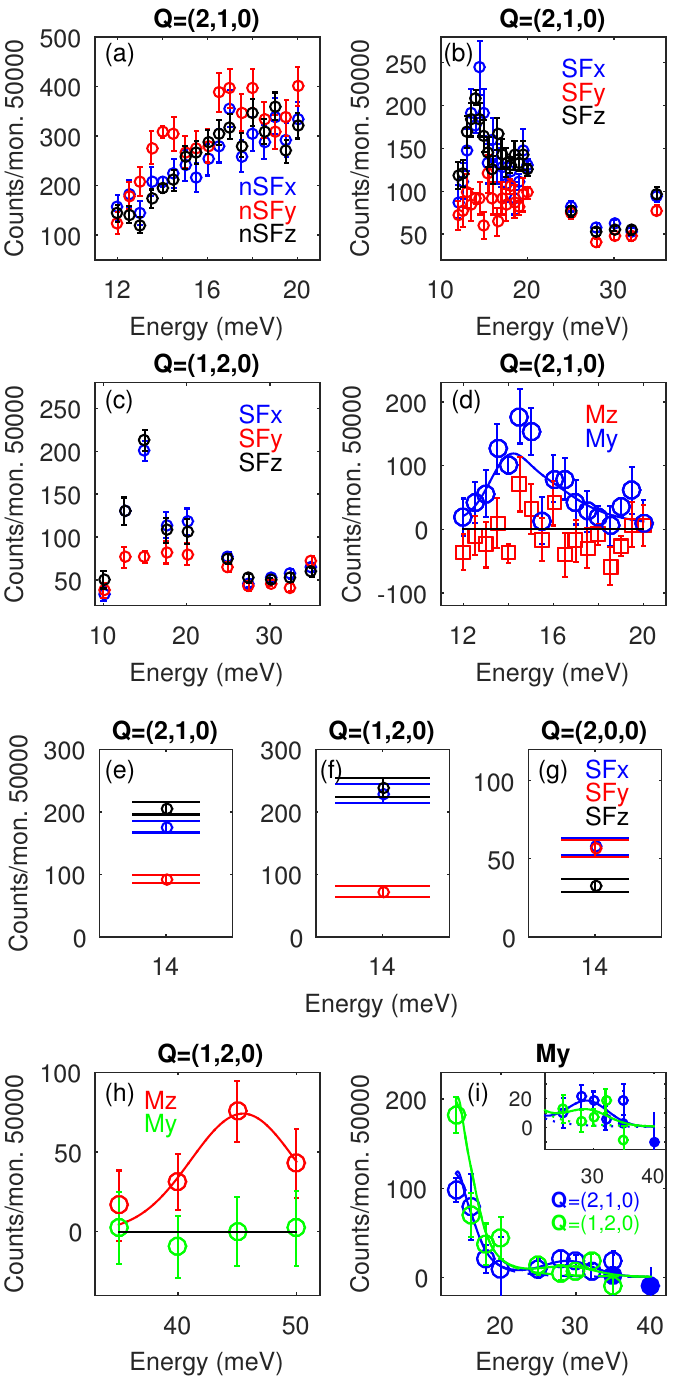}%
 \caption{\label{Modes}(color online) Energy scans at various \textbf Q values with polarization analysis. Raw data of different (a) nSF,  and (b,c,e-g) SF  channels with the polarization parallel to \textbf x (\textbf y),(\textbf z) in blue (red), (black) at different scattering vectors. (d,h,i) show the results of the polarization analysis with Eq. (1), in red M$_z$, and  M$_y$ in blue (green) at \textbf Q=(2,1,0) (\textbf Q=(1,2,0)).  In (d) the blue line depicts the modeled asymmetric shape resulting from the folding of the magnon dispersion with the resolution function, see Ref.~\cite{Kunkemoeller2015},  which is also used for the modeling in (i), but scaled with the expected correction factors for geometry and twinning fractions. The inset in (i) shows a zoom into the data.}
 \end{figure}

With the alignment of the antiferromagnetic moment along $b$ one expects non-degenerate transverse magnon modes corresponding to polarization along
$a$ or $c$ \cite{Qureshi2012}. With our previous unpolarized inelastic neutron scattering experiments \cite{Kunkemoeller2015} only the $a$ polarized mode could be clearly identified. The gap of the $c$ polarized
transverse mode could either be hidden in the shoulder of the in-plane signal, which inevitably arises from the folding of the instrument resolution with
the steep spin-wave dispersion, or appear at higher energies where weak signals were detected \cite{Kunkemoeller2015}. With the new
polarized experiment the in-plane and $c$ polarized transverse magnons can be easily separated in the polarization and it was the first
aim to search for the $c$ polarized mode.

Figure~\ref{Modes}(a-c) shows energy scans through the anisotropy gap at the antiferromagnetic zone center. The nSF channels shown in Fig.~\ref{Modes}(a) have a larger background than the SF channels [Fig.~\ref{Modes}~(b-c)] and are thus less informative but confirm the main conclusions.
The $c$ polarized mode must entirely contribute
to the SFy and nSFz channels with a complete geometry factor.
Spin-wave calculations show that the two non-degenerate transverse modes exhibit an intrinsic signal strength
inversely proportional to their zone-center energy \cite{Qureshi2012}. Therefore, the $c$ polarized mode can be excluded in the asymmetric peak of the in-plane transverse mode. These scans
confirm the in-plane character of the signal in the range 14 to 20~meV, see Fig.~\ref{Modes}(d). A special effort was laid on the analysis of the signal maximum at the in-plane magnon gap, E=14~meV, see Fig.~\ref{Modes}(e,f). With the partially detwinned  crystal one expects this signal from the transverse magnon to be reduced by a factor 0.6 for $\textbf Q=(2,1,0)$, which agrees with the measured ratio of the intensities at 14~meV between $\textbf Q=(2,1,0)$ and $\textbf Q=(1,2,0)$ of 0.55(10).
We also measured the backfolded mode at the ferromagnetic zone-center, which according to the spin-wave calculations, see below, exhibits a
$c$ polarization. Indeed such a $c$ polarized signal can be determined at $Q=(2,0,0)$ and E=14~meV [Fig.~\ref{Modes}(g)] by counting for a very long time (45~min for each data point).

In view of the recent observation of longitudinal magnetic excitations in CRO \cite{Jain} it seemed worthwhile to search for such modes first at the antiferromagnetic
zone centers (2,1,0) and (1,2,0).
In Fig.~\ref{Modes}(i) the fitted background from the in-plane polarized transverse mode from Fig.~\ref{Modes}(d) is used to separate a weak signal peaking at 29~meV. The signals for the two different \textbf Q values are assumed to posses the same shape, but are scaled with the expected factors for geometry and twinning. Around 30~meV the stronger signal for $\textbf Q=(2,1,0)$ compared to $\textbf Q=(1,2,0)$ thus points to a longitudinal mode in the sense that it is polarized parallel
to the static moment. The energy of this mode corresponds fairly well to the double of the gap of the in-plane transverse branch and the strength is below 10~\% of that signal. Therefore it seems most likely that this longitudinal signal completely stems from the two-magnon excitation, which is expected to appear in the longitudinal polarization channel \cite{Heilmann1981}.  A similar discussion has also been initiated for the parent material of FeAs-based superconductors
where again the two-magnon explanation seems most likely \cite{Qureshi2012,Wang2013,Fidrysiak2016,Waser2016}. The 30~meV longitudinal signal is thus not anomalous.

The $c$ polarized antiferromagnetic zone-center mode is detected at higher energy, where the unpolarized experiment found some evidence for additional scattering \cite{Kunkemoeller2015}, but where the signal strength is expected to be rather small \cite{Qureshi2012}. In order to cover the energies of the order of 40~meV we needed to use a larger value of $k_f=4.1$~\AA$^{-1}$, which also allows one
to avoid the contamination appearing at E=44~meV for the standard value of $k_f=2.662$~\AA$^{-1}$.
The magnetic signals polarized along $\textbf z\simeq c$ and along \textbf y at large energies are shown in Fig.~\ref{Modes}(h) and show that the zone-center $c$-polarized magnon
possesses a large energy of 45.5(1.5)~meV. Its signal strength is in rough agreement with the signal strength of the in-plane transverse mode with the energy being enhanced by a factor three. Our conclusion of $c$ polarized modes at the ferromagnetic and antiferromagnetic zone centers agrees with the interpretation of similar polarized neutron scattering experiments performed with twinned crystals in a different scattering geometry \cite{Jain}.

This large $c$ anisotropy is remarkable, as it strongly deforms the magnon dispersion in CRO with respect to a simple isotropic model. The splitting of the
two antiferromagnetic zone-center magnons, 14 and 45.5~meV, is larger than the dispersion of the in-plane branch to the zone-boundary. As will be discussed below,
the entire dispersion of transverse branches is, nevertheless, well described with the $S=1$ spin-wave model using the Holstein-Primakoff transformation. The in-plane and $c$-polarized
branches exchange their character: while the $c$-polarized mode is the high-energy mode at the antiferromagnetic zone center, it appears at the
lower energy of the in-plane transverse magnon at the ferromagnetic zone center, in accordance to the data shown in Fig.~\ref{Modes}(g).
The in-plane polarized branch thus starts at 14~meV at the antiferromagnetic zone center, exceeds to the zone boundary at 41~meV and then continues to stiffen till
45.5~meV at the ferromagnetic zone center. The out-of-plane polarized branch just exhibits the opposite dispersion. By comparing the intensities at the ferromagnetic zone centers $\textbf Q=(2,0,0)$ and $\textbf Q=(0,2,0)$ with an energy transfer of 45~meV we may confirm the in-plane transverse character of this high-energy mode appearing at
the ferromagnetic zone center. The expected ratio of the signals taking into account the different twinning fractions and geometry factors amounts to 1:2.5 for a  transverse and to 2.5:1 for a longitudinal magnon. We obtain a ratio of 0.1(2) indicating the transverse in-plane character. Evidence for a longitudinal high energy branch could not be obtained in our experiment \cite{Jain,note}.


\subsection{Spin-wave calculations of the magnon dispersion}

  \begin{figure}
   \includegraphics[width=0.9\columnwidth]{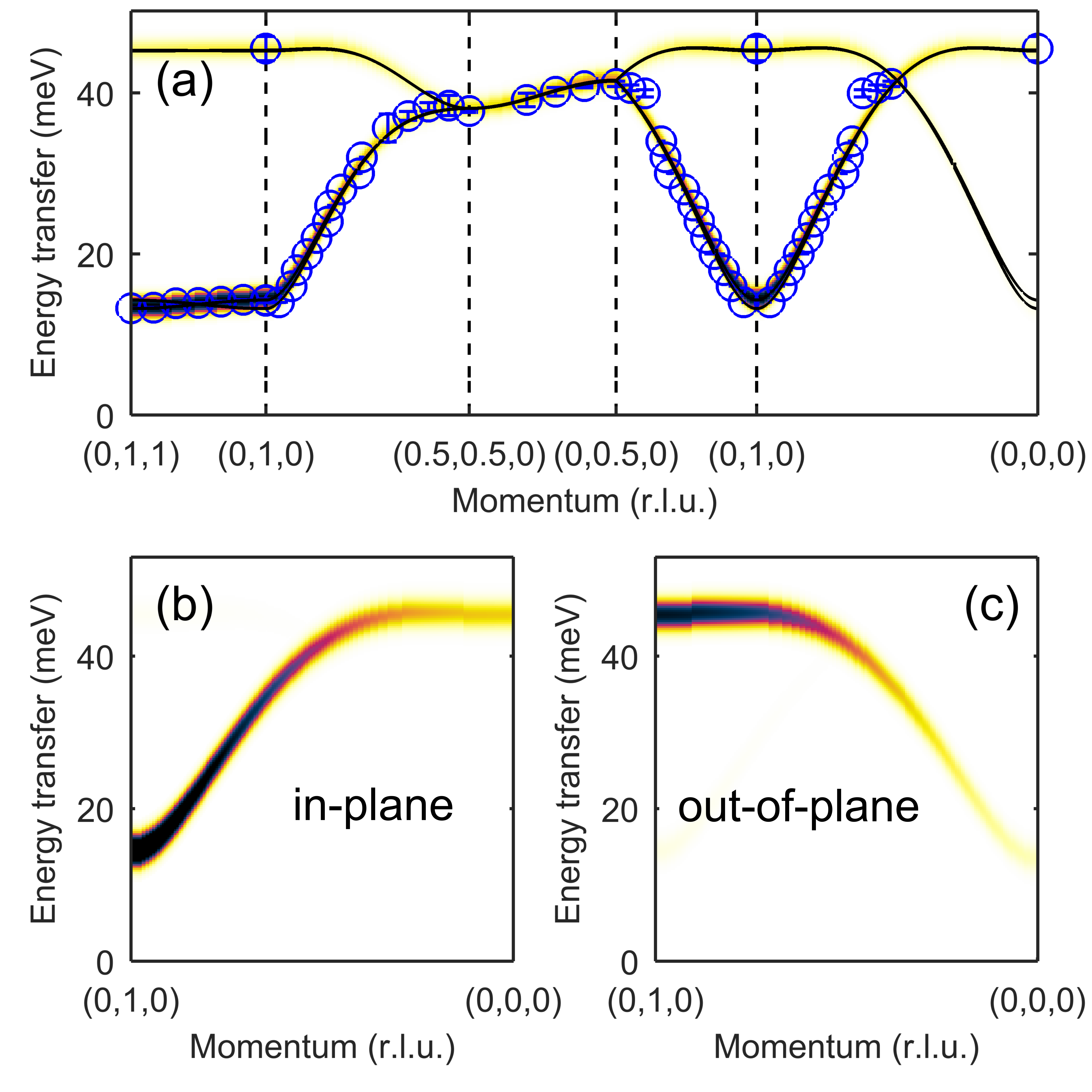}
   \caption{\label{SpinW} Magnon dispersion calculated with the SpinW program using the values given in the text. The blue circles denote the fit values of the dispersion obtained in this and the previous \cite{Kunkemoeller2015} neutron scattering studies. The black lines are the calculated dispersion and the color code denote the calculated neutron scattering intensity of the convoluted spectra. In each panel black stands for maximum intensity and white for none. In (b,c) only the in-plane- and out-of-plane polarized transverse modes are shown, respectively, illustrating the opposite dispersion of these branches.}
   \end{figure}

The anisotropy gap of the $c$-polarized transverse magnon of 45.5~meV leads to the uncommon feature in the magnon dispersion,
that the transverse in-plane branch continues to increase in energy between the zone boundary and a ferromagnetic zone center. This peculiarity can,
however, be well described with a rather conventional model.

The Hamiltonian, which is used for the description of the magnon dispersion is given by:

\begin{equation}
H =
\sum_{\mathbf{i,j}}\mathrm{J}_{i,j}\mathbf{S_i}\cdot\mathbf{S_{j}}+\gamma
\sum_i  (S^x_{i})^2+\epsilon\sum_i  (S^z_{i})^2.
\end{equation}

The sum runs over pairs of magnetic ions, so that each pair or bond appears twice and $S$ is set to 0.67 following the experimental results in Ref.~\cite{Braden1998}. The spin-wave calculations using the SpinW program and the parameters $J=5.6$~meV, $J_{na,b}=0.6$~meV, $J_c=-0.03$~meV, $\gamma=1.4$~meV and $\epsilon=24.5$~meV, give a good description of the magnon dispersion obtained in this and the previous \cite{Kunkemoeller2015} neutron scattering studies [Fig.~\ref{SpinW}(a)]. Panels Fig.~\ref{SpinW}(b,c) show the in-plane and c-polarized branches, respectively, starting at an antiferromagnetic zone center and proceeding to a ferromagnetic one. In this figure the color code denotes maximum neutron scattering intensity of the convoluted spectra with black color, and zero intensity with white. The energy resolution is set to 1~meV. The spin-wave calculation thus perfectly describes not only the energy dispersion but also the polarization of the magnon branches. Note that this rather uncommon dispersion is described with strong single-ion parameters arising from SOC. These strong anisotropy terms interfere with the Heisenberg interaction parameters in contrast to models invoking only weak anisotropy. Therefore, the $J$ parameter differs from that
obtained by fitting only the lower part of the dispersion with a small uniaxial anisotropy \cite{Kunkemoeller2015}. Figure~\ref{SpinW}(a) shows that the non-sinusoidal parts of the dispersion are perfectly described, clearly better than without the strong anisotropy terms. It is, however, important to note that the model is
not unique. It is possible to obtain similar fitting by partially reducing the single-ion anisotropy and by inducing an anisotropic nearest-neighbor interaction.

%

The coupling parameter J$_c$ acting between neighboring layers splits the magnon modes into two by introducing a finite dispersion perpendicular to the planes. Whether the lower or upper mode is seen at $\textbf Q=(0,1,0)$ is determined by the sign of $J_c$. With the partially detwinned crystals (in particular with the
experiment on 1TiB) it is possible to determine the sign of this interaction.
Here it is chosen to couple spins at (0,0,0) and (0,0.5,0.5). The dominant $b$ components of these spins are parallel for the $A$-centering and antiparallel for the $B$-centering. If $J_c$ is positive (negative), corresponding to an antiferromagnetic (ferromagnetic) coupling, the upper (lower) mode is seen at $\textbf Q=(0,1,0)$. The experimental data reveal that the upper mode is seen at $\textbf Q=(0,1,0)$, so the coupling is ferromagnetic, stabilizing the $A$-centered phase. This coupling contradicts the observation of a $B$-centered structure as the main magnetic scheme, see subsection III.B. CRO thus exhibits the uncommon situation that the minimum
magnon energy does not occur at the magnetic Bragg peaks. This observation can be attributed to an anisotropic $J_c$ which differs for the
$a$ and $b$ spin components.

\subsection{Density functional calculations of magnetic structure and interaction}

The linearized augmented plane wave (LAPW) method as implemented in the wien2k package\cite{Blaha2001} was used for the DFT calculations. The exchange correlation potential for the generalized gradient approximation (GGA) was chosen to be in the form proposed in Ref.~\cite{Perdew1996}.  The SOC was treated in a second variational way. The parameter of the plane-wave expansion was chosen to be $R_{MT}K_{max}=7$, where $R_{MT}$ is the smallest atomic sphere radii ($R_{MT}^{Ca}=2.15$~a.u., $R_{MT}^{Ru}=1.99$~a.u., $R_{MT}^{O}=1.71$~a.u.) and $K_{max}$ the plane-wave cutoff. We used a mesh consisting of 800~$k-$points. The on-site Hubbard repulsion and intra-atomic Hund's exchange were taken to be $U=3$~eV and $J_H=0.7$~eV\cite{Lee2006a,Streltsov2012a} in the GGA+U\cite{Anisimov1997} and GGA+U+SOC calculations.

The exchange parameters were recalculated via total energies of three different collinear magnetic configurations (a ferromagnetic and two, in the $c$ direction differently stacked, antiferromagnetic configurations) using the GGA+U approach. We found that $J=4.9$~meV, which agrees with the results of the spin-wave calculations.

In order to estimate the single-ion anisotropy we add SOC to the calculation scheme and computed the energies of two configurations, where all spins are either directed along $c$ or lie in the $a,b$~plane. The antiferromagnetic order was assumed in these calculations. The lowest total energy corresponds to the configuration, where all spins are directed predominantly along the $b$~axis. The single-ion anisotropy is found to be very large, $\epsilon=20.1$~meV, again in good agreement with experiment. Very recently microscopic magnetic parameters were calculated
for similar $U$ and $J_H$ values finding a nearest-neighbor interaction of 3-6\,meV and a large single ion anisotropy in
agreement with our results \cite{pavarini}.

The spin moment on the Ru ions was calculated to be $M_s^{Ru} =1.27~\mu_B$ (i.e. $S=0.63$), while the orbital moment is $M_l^{Ru} =0.13~\mu_B$. We also find a sizable magnetic moment on the apical O $M_s^{O}=0.11~\mu_B$, while planar ligands stay nonmagnetic. The total moment, $M_s^{Ru}+2M_s^{O}-M_l^{Ru}$, agrees well with the powder neutron diffraction result, and the small value of the orbital moment indicates that CRO is not close to a $j=0$ state driven by SOC as it is also deduced from the calculations in reference \cite{pavarini}.

\subsection{The influence of Ti substitution}
  \begin{figure}
  \includegraphics[width=0.9\columnwidth]{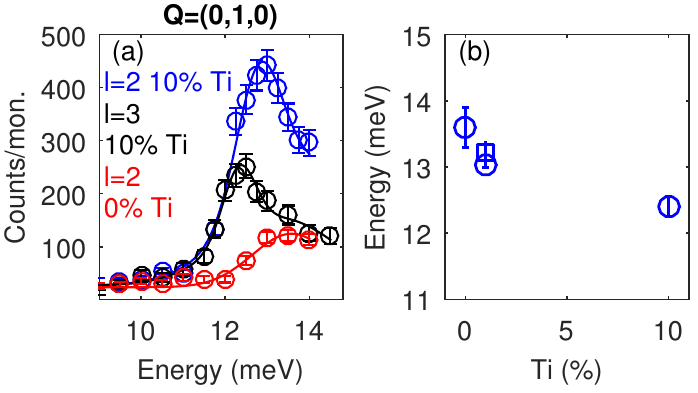}%
  \caption{\label{Ti}(color online) Influence of Ti substitution on the anisotropy gap. (a) shows energy scans through the anisotropy gap at the antiferromagnetic zone center using the crystal 10Ti, l=2 (blue) and l=3 (black) and 0Ti, l=2 (red). In (b) the energy of the lower zone-center mode of crystals with different Ti content is shown, the data for 1~\% Ti are from Ref.~\cite{Kunkemoeller2015}.}
  \end{figure}

In Fig.~\ref{Ti} the influence of different Ti substitutions on the anisotropy gap is addressed by comparing constant \textbf Q scans
across the in-plane gap at the antiferromagnetic zone center.
The modes are slightly split due to finite inter-layer interaction, see subsection III.D. Even $l$ corresponds to the higher and odd to the lower modes, respectively. The lower zone-center magnon energy is displayed as function of Ti doping in Fig.~\ref{Ti}(b). The gap clearly diminishes with increasing Ti content. Since the non-magnetic Ti dilutes the magnetic lattice
a general softening can be expected, as it is visible in the anisotropy gap, which in first view corresponds to the square root of exchange and anisotropy
energies. In addition, Ti also perturbs the lifting of orbital degeneracy, as it is shown in Fig.~\ref{RuOOO} and subsection III. A. In consequence also the
single-ion anisotropy will be reduced with increasing Ti content. The impact of a small Ti content of the order of one percent, which is used to
favor the synthesis of large crystals, can however, safely be neglected. 10Ti, which is essentially untwinned, shows a $l$ dependence of the anisotropy gap [Fig:~\ref{Ti}(a)] consistent with the previous study \cite{Kunkemoeller2015} and the discussion given above.

\subsection{Additional modes and magnetic polarization of oxygen}

 \begin{figure}
 \includegraphics[width=0.9\columnwidth]{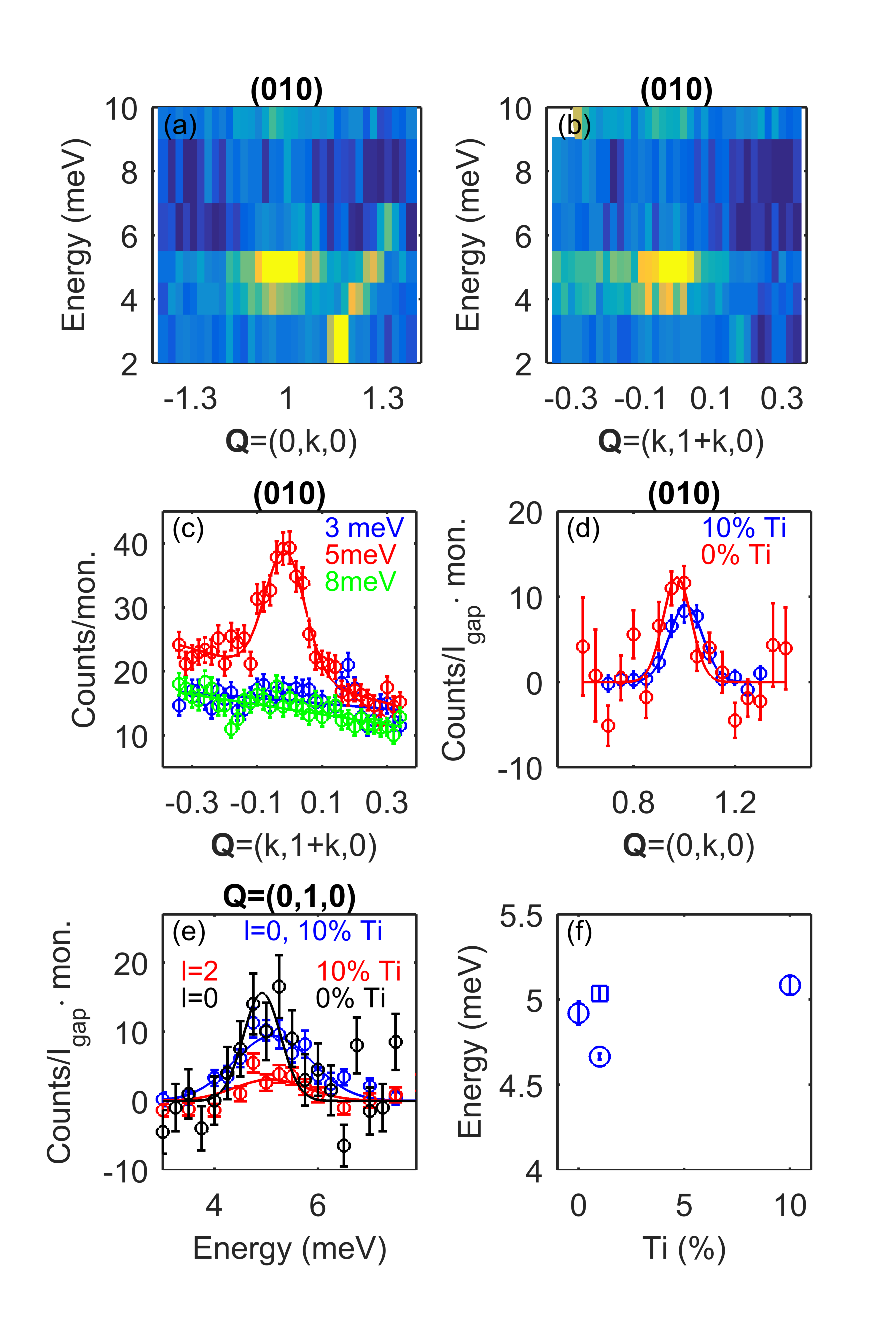}%
 \caption{\label{OMode2}(color online) Additional magnon branch near 5~meV. (a,b) show scattering maps of energy versus \textbf Q. (c) presents some characteristic scans included in (b) for an energy transfer of 3~meV (5meV), (8~meV) in blue (red), (green). The data in (a-c) are taken from Ref. \cite{Kunkemoeller2015}. (d) shows \textbf Q scans with an energy transfer of 5~meV for 10Ti (blue) and 0Ti (red). The intensities are scaled with the maximum intensities of the scans in Fig.~\ref{Ti}. Scaled in the same way are the energy scans in (e) at $\textbf Q=(1,0,l)$ using the crystals 10Ti, $l=0$ (blue), $l=2$ (red) and 0Ti, $l=2$ (black). (f) presents the energy of this additional mode at \textbf Q=(0,1,0) for crystals containing different amounts of Ti.}
 \end{figure}

 \begin{figure}
 \includegraphics[width=0.9\columnwidth]{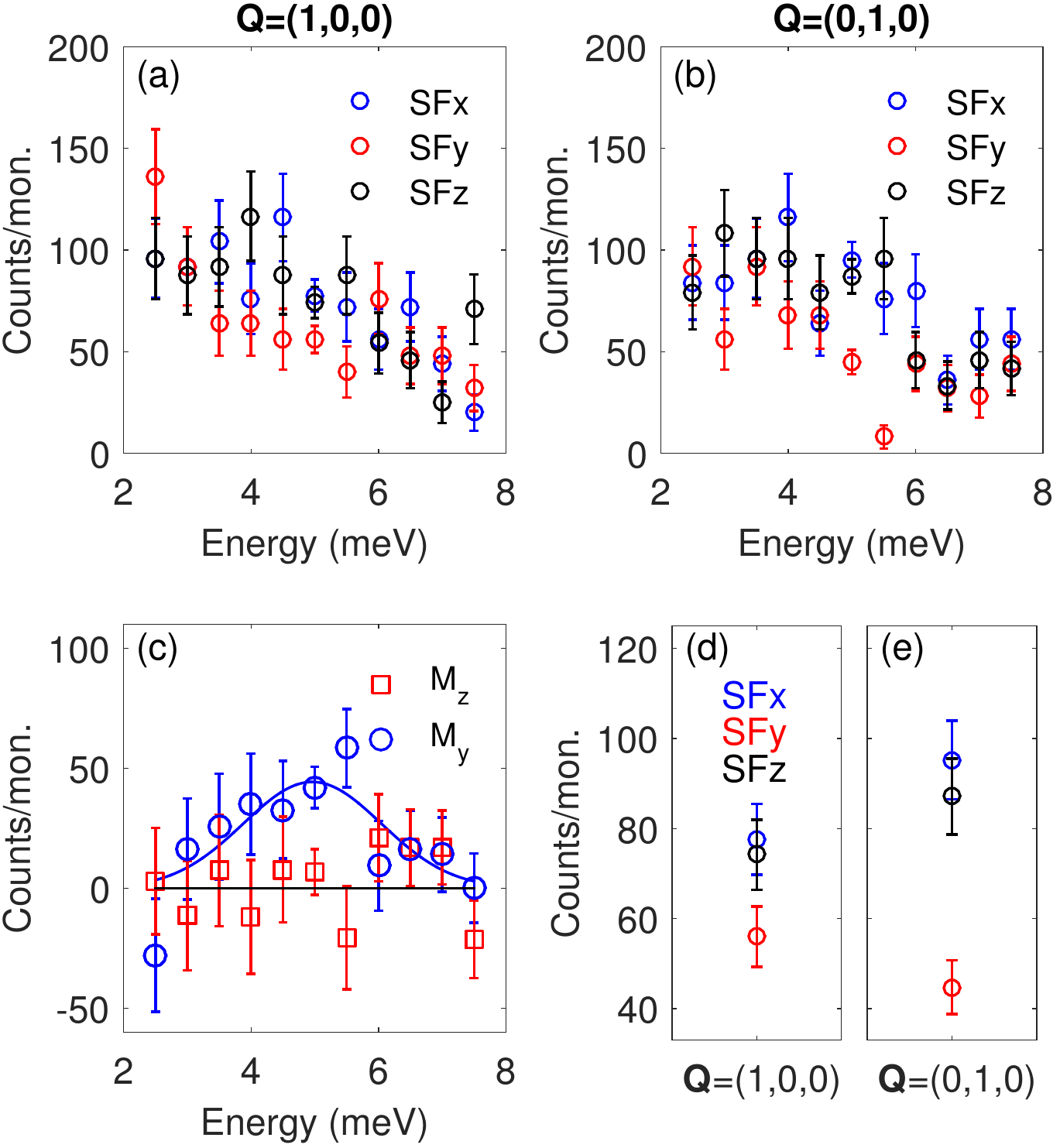}%
 \caption{\label{OModePol}(color online)  Polarization analysis of the additional mode. SF intensities of energy scans at (a,d) $\textbf Q=(1,0,0)$, and (b,e) $\textbf Q=(0,1,0)$ for the SFx (SFy), (SFz) chanel in blue (red), (black). (c) shows the results of the polarization analysis with Eq. (1), in red M$_z$ and in blue M$_y$. (d,e) present only the 5~meV data of the two \textbf Q positions, which are measured with higher statistics. }
 \end{figure}

The previous unpolarized experiments gave evidence for an additional scattering at the lower energy of 5\,meV that could not be
explained by the magnon dispersion expected for a simple square-lattice antiferromagnet. The data, which are presented in the mappings shown in
Fig.~\ref{OMode2}(a,b), were obtained on the thermal TAS IN8 \cite{Kunkemoeller2015}. Panel (c) shows some characteristic scans included in (b).
The 5~meV signal is much weaker than the in-plane transverse mode in particular when considering the by a factor
three reduced energy. Furthermore, it is not possible to connect this low-energy feature with a dispersing branch with a slope comparable to that of the transverse branch. Instead the intensity of this feature seems to be rather localized in \textbf Q space and to exhibit a flat dispersion.  The signal could also be followed along the vertical direction without any indication for finite dispersion.

In Ref.~\cite{Kunkemoeller2015} it was speculated that this extra mode could arise from the orbital disorder induced by the Ti substitution, a scenario which can now be ruled out by the comparison of Ti-containing and Ti-free CRO.
Fig.~\ref{OMode2}(d-e) show \textbf Q and energy-scans across the 5~meV signal on samples with 0 and 10\% of Ti. Most interestingly there seems to
be no difference in the strength of the low-energy signal when normalized to that of the in-plane gap mode at $\sim$14~meV. The additional signal is thus intrinsic to the magnetic order in CRO, and it is not induced by disorder, which must be significantly enhanced in Ti10.
The Ti content furthermore does not change the energy of the additional scattering, see the fitted energy maxima displayed in Fig.~\ref{OMode2}(f). The minor variation between measurements on the same concentration on different spectrometers can be attributed to the energy calibration.

The polarization of the additional mode was determined with polarized neutron scattering following the same procedures as those used to determine the character of the dispersion at higher energies. The polarization
analysis directly shows that the additional excitation possesses an in-plane character, because it is not observed in the SFy channel (or in $M_z$), see Fig.~\ref{OModePol}. Comparing the intensities at $\textbf Q=(1,0,0)$ and $\textbf Q=(0,1,0)$ one may furthermore deduce that the additional mode is $a$~polarized, which is transverse with respect to the static moment. The expected ratio of the intensities at 5~meV between $\textbf Q=(0,1,0)$ and $\textbf Q=(1,0,0)$ based on the twinning fractions and the geometry factor is 1:2.5 for an in-plane transverse mode and 2.5:1 for an in-plane longitudinal mode, respectively. The measured ratio is 0.25(21) characterizing this mode as an in-plane transverse mode.

The simple antiferromagnetic square lattice exhibits a magnon dispersion consisting of two transverse branches, which were both observed at larger energy
in CRO, see above. In order to explain the additional mode the model needs to be extended, and the weakness and localization of the signal point to a smaller
moment with smaller coupling.

The GGA+U calculations on CRO reveal a small magnetic moment situated at the two apical oxygens of 0.11\,$\mu_B$ each. Such a polarization of the oxygen
results from the strong hybridization between Ru $4d$ and O $2p$ orbitals. DFT calculations on several ferromagnetic or nearly ferromagnetic ruthenates find
a sizable amount of magnetic moment on O, which in total sum up to about 30~\% of the entire magnetization \cite{Mazin1997}. Experimentally this sizable magnetization of the
O states has been observed by polarized neutron diffraction on Ca$_{1.5}$Sr$_{0.5}$RuO$_4$ where indeed 33~\% of the magnetization stems from O states \cite{Gukasov2002}. CRO, however, exhibits antiferromagnetic order, in which the O bridging two antiparallel Ru moments cannot be polarized, see Fig.~\ref{explanations}(a). Therefore, there is no moment on the in-plane
O sites, which is also confirmed by the DFT calculations of CRO. However, the apical O is connected only to a single Ru site and can be polarized.

 There are two principal contributions to the exchange between Ru and O moments due to hopping between O 2$p$ and Ru 4$d$ orbitals in the scenario where the Ru $d_{xy}$ orbital is doubly occupied and the $d_{xz}$ and $d_{yz}$ singly occupied, see e.g. Ref.~\cite{Khomskii2014}. One is a hopping from the $p_x$ and $p_y$ to the $d_{xz}$ and $d_{yz}$ orbitals, respectively. There, because of Pauli exclusion principle, only antiparallel spins can hop, leaving parallel spins in the now singly occupied O $p$-orbitals. In total this yields a ferromagnetic coupling:
  \begin{eqnarray}
  \label{t2U}
  J_{Ru-O}^{t_{2g}-p} \sim  \frac{-2t_{pd \pi}^2}{\varepsilon_{t_{2g}} - \varepsilon_p},
  \end{eqnarray}
  where $t_{pd \pi}$ is the hopping integral between $d_{xz}$ and $p_x$, and $d_{yz}$ and $p_y$ orbitals of Ru and oxygen. $\varepsilon_{t_{2g}}$ and $\varepsilon_p$ are the centers of the Ru $t_{2g}$ and apical O $p$ bands. The factor two in Eq. (\ref{t2U}) reflects the fact that there are two processes (virtual hoppings from $p_x$ to $d_{zx}$ and from $p_y$ to $d_{yz}$ orbitals), contributing to this ferromagnetic Ru-O exchange.

 The second contribution to the Ru-O exchange comes from the hopping from the $p_z$ orbital of O to the empty $d_{3z^2-r^2}$ orbital of Ru. Because of Hund's rule, predominantly parallel spins hop, leaving anitparallel spins on the O $p_z$ orbital. This yields an antiferromagnetic coupling between Ru and O moments:
 \begin{eqnarray}
 \label{t2JD2}
 J_{Ru-O}^{e_g-p} \sim  \frac {t_{pd \sigma}^2}{\varepsilon_{e_{g}} - \varepsilon_p } \frac {J_H}{\varepsilon_{e_{g}}-\varepsilon_p}.
 \end{eqnarray}
   Here $t_{pd \sigma}$ is the hopping matrix element between a $p_z$ orbital of O and an empty $d_{z^2-r^2}$ orbital of Ru, cf. Ref. \cite{Khomskii2014}.

The $\sigma$ hopping is larger than the $\pi$ one $(|t_{pd \sigma}| \sim 2.2 |t_{pd \pi}|$ )\cite{ Harrison1999}, but the antiferromagnetic exchange (Eq. \ref{t2JD2}) occurs through the $e_g$ band, which lies much higher in energy than $t_{2g}$. Our GGA+U calculations show that $\varepsilon_{t_{2g}} - \varepsilon_p  \sim 1$ eV, while $\varepsilon_{e_{g}} - \varepsilon_p \sim$4.7~eV. There is  only one such antiferromagnetic exchange path, in contrast to two for ferromagnetic exchanges (Eq. \ref{t2U}) and Hund’s exchange on the Ru, $J_H \sim 0.7$ eV, also slightly reduces this exchange (Eq. \ref{t2JD2}). Therefore, one should expect ferromagnetic coupling to dominate, so that the moment of oxygen would be parallel to the moment of neighboring Ru. Our DFT calculations support this conclusion.

 \begin{figure}
 \includegraphics[width=0.9\columnwidth]{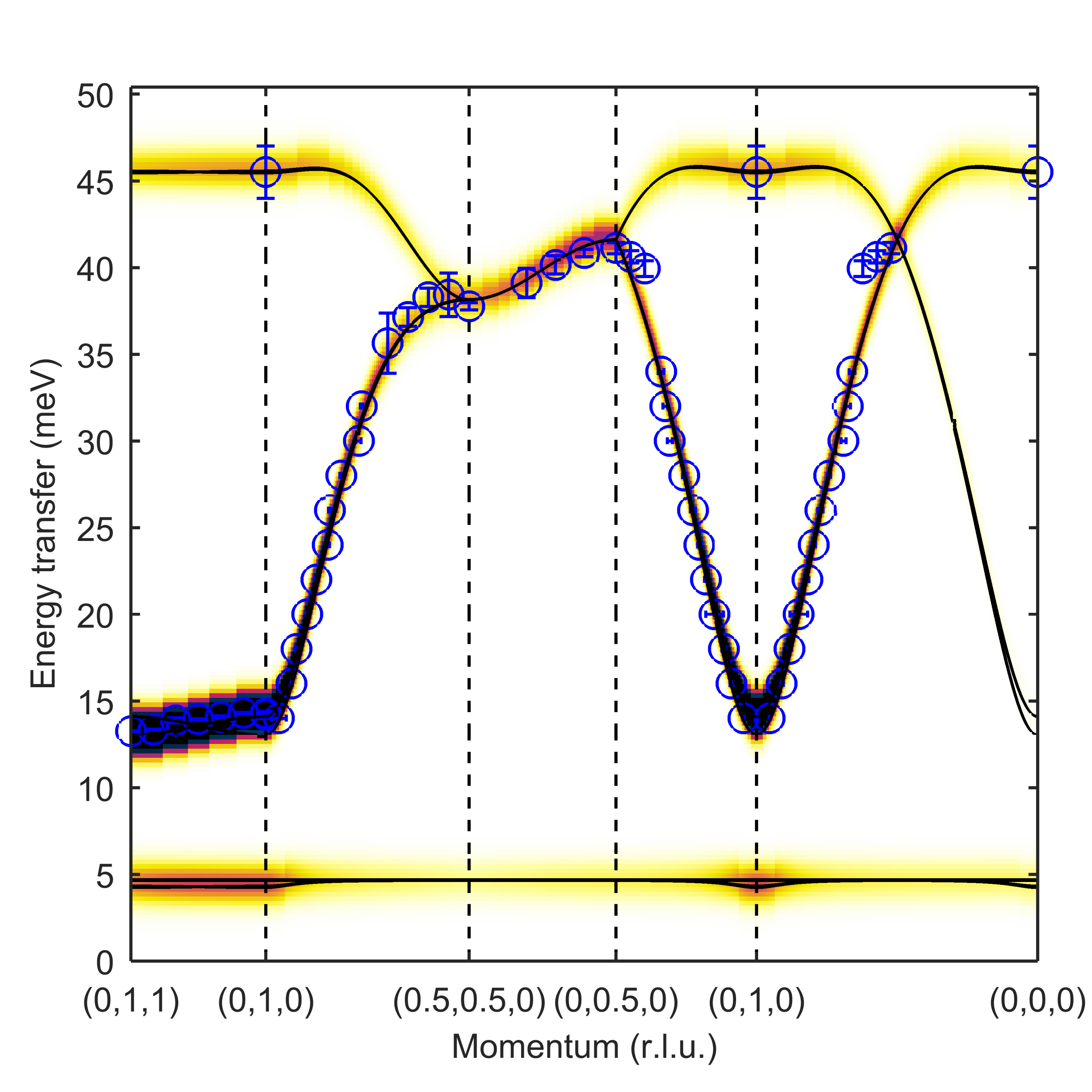}%
 \caption{\label{OModeDisp}(color online) Magnon dispersion calculated with the SpinW program including the additional mode. The labeling corresponds to the one in Fig.~\ref{SpinW}.  }
 \end{figure}

We have extended the spin-wave model by a small moment on the apical oxygens, 0.11~$\mu_B$, coupled ferromagnetically to the next Ru moment, see Fig.~\ref{explanations}(a). This extension necessitates to reduce the single-ion anisotropy parameter $\gamma$ in order to keep the anisotropy gap of the in-plane transverse mode at the correct energy. The additional mode cannot be obtained at an energy near 5~meV at the antiferromagnetic zone center without introducing an anisotropy in terms of a single ion anisotropy for the magnetic moment on the O sites, or an anisotropic coupling parameter. Because the single ion anisotropy stems from SOC which is small in O, the first scenario seems to be less likely. A better choice is an anisotropic coupling, because the Ru with its larger SOC is also involved in this process. Therefore we used $J_{RuO}=(-1.5,-3.5,-1.5)$~meV to fit the experimental results. With this anisotropic coupling, $\gamma$ has to be reduced to 0.5~meV to keep the in-plane transverse mode at the experimental value.
 With this extension we can perfectly describe the additional feature, see Fig.~\ref{OModeDisp}. The flat dispersion of the additional branch results from the lack of coupling between two O moments and perfectly describes the experimental finding. Also the localization of the signal strength at the antiferromagnetic zone center is well reproduced by this model. We can therefore conclude that the additional moment, situated on the apical O site and ferromagnetically coupled to the Ru, can explain the additional low-energy magnetic signal in CRO. We also tried to reproduce the data with the O moments being antiferromagnetically coupled to Ru, but such a scenario cannot explain the location of the scattering in Q-space.

The significant polarization of the O site is another consequence of the strong hybridization of the Ru $4d$ and O $p$ orbitals resulting in large hopping parameters, which can already be deduced from the closeness of the metallic phase of CRO. It furthermore yields a convincing explanation of the reduced
moment, because the in-plane oxygen ion, which should carry a similar amount of charge carriers, cannot be magnetically polarized in the antiferromagnetic structure reducing the ordered moment from the simple $S=1$ expectation. The ordered moment may thus also oscillate in length by transferring
moment between Ru and in-plane O, which can give rise to a longitudinal mode, as an alternative explanation for the observation in reference \cite{Jain}.

\section{Conclusions}

Various neutron scattering experiments on partially untwinned crystals of Ca$_2$Ru$_x$Ti$_{1-x}$O$_4$ were performed to
determine the magnetic structure and magnon dispersion in the insulating state of CRO. Most remarkable is the
large magnetic anisotropy, which results in a splitting of the two transverse zone-center modes that exceeds the full dispersion.
The entire dispersion of transverse magnon branches, however, is perfectly described by a conventional spin-wave model,
in which the strong impact of the SOC is reflected by large anisotropy parameters. DFT calculations within
the GGA+U approximation yield an orbital moment of only 0.13\,$\mu_B$, which indicates that CRO is not close to a $j=0$ state, which can be seen in the discrepancy of the experimental magnon dispersion with the $j=0$ calculation \cite{Akbari2014}.
The calculated magnetic interaction and anisotropy parameters reasonably well agree with those obtained by fitting the dispersion,
and also the ordered moment is correctly reproduced in the DFT calculation.

The DFT study finds a sizable amount of magnetic moment on the apical oxygen reminiscent of previous reports on ferromagnetic ruthenates.
This additional moment explains an additional signal appearing in the neutron scattering experiments at lower energy and limited
to the antiferromagnetic zone center. Extending the spin-wave model to the weakly ferromagnetic coupled oxygen moments describes the flat dispersion and
the limited appearance in \textbf Q space of this signal.

The spin-wave dispersion in CRO is thus dominated by the impact of strong SOC and by the presence of magnetic moments
on the oxygen sites.

\section{Acknowledgments}
This work was supported by the Deutsche Forschungsgemeinschaft through CRC 1238 Projects No. A02 and No. B04 and by the Russian Foundation of the Basic Research via Program 16-02-00451, Russian president council on science through MD-916.2017.2, FASO (theme ``electron'' 01201463326) and MON (project 236).

\appendix
\section{Crystal structure determination of Ca$_2$Ru$_x$Ti$_{1-x}$O$_4$}
Complete crystal structure analyses were performed with an  X8-APEX by Bruker AXS single-crystal diffractometer with a goniometer
in kappa-geometry and x-ray radiation from a molybdenum anode with a wavelength of $\lambda=0.71073$~\AA. The distance between the sample and the detector
was set to 50~mm. Structure refinements were carried out using Jana2006 \cite{Petricek}. A type I extinction correction was applied during the refinements and the data were corrected for absorption. The thermal parameters for Ru1 and Ti1 were constrained to be equal, and the total occupancy of this site was fixed to 1. The results of the structural refinements are given in Table \ref{tab:ref}. Table \ref{tab:char} presents further characteristics of the crystal structure as well as the metal-insulator- and antiferromagnetic transition temperatures. $T_N$ is determined from the magnetization curves presented in Fig.~\ref{Magnetization}(a) by finding the zero point of the second derivation. $T_{MIT}$ is determined in the same way from the resistance curves presented in this figure.

 \begin{table*}[htbp]
\caption{Structural data of the single crystal x-ray diffraction analysis.}
\vspace{5mm}
           \centering
         \begin{tabular}{l  llllll}

         \hline \hline
                   Ti-content    &   {10\%} &   {10\%} &
     {1\%} &   {1\%} & 0\%   & 0\% \\
     {Temperature} & 100K  & 293K  & 100K  & 293K  &   {100K} &
     {293K} \\

   \hline
                   wR(all,F) &   {0.022} &   {0.0429} &
     {0.0521} &   {0.048} & 0.0549 & 0.0262
   \\ \hline

                   Ru1, Ti1   &       &       &       &       &       &  \\
                   x     &   {0} &   {0} &
     {0} &   {0} & 0     & 0 \\
                   y     &   {0} &   {0} &
     {0} &   {0} & 0     & 0 \\
                   z     &   {0} &   {0} &
     {0} &   {0} & 0     & 0 \\
                   U$_{11}$ & 0.00119(3) & 0.00284(3) & 0.00197(5) & 0.00348(5) &
     {0.00295(4)} &   {0.00368(4)} \\
                   U$_{22}$ & 0.00164(4) & 0.00330(4) & 0.00245(5) & 0.00368(5) &
     {0.00373(5)} &   {0.00409(5)} \\
                   U$_{33}$ & 0.00151(4) & 0.00378(3) & 0.00262(4) & 0.00390(4) &
     {0.00351(5)} &   {0.00403(4)} \\
                   U$_{12}$ & -0.00003(3) & -0.000019(15) & 0.00000(4) & -0.00001(2) &
     {0.00001(3)} &   {-0.000015(12)} \\
                   U$_{13}$ & -0.00009(2) & -0.000172(13) & -0.00005(3) & -0.00019(2) &
     {-0.00011(4)} &   {-0.000187(12)} \\
                   U$_{23}$ & -0.00011(2) & -0.00025(2) & -0.00012(3) & -0.00021(2) &
     {-0.00006(3)} &   {-0.000228(14)} \\
                   occupancy Ru & 0.870(2) & 0.866(2) & 0.969(3) & 0.978(3) &
     {0.9768(18)} &   {0.9859(17)} \\
    occupancy Ti & 0.130(2) & 0.134(2) & 0.031(3) & 0.022(3) &
        - &   - \\ \hline
                   Ca1   &       &       &       &       &       &  \\
                   x     & 0.00753(3) & 0.01019(2) & 0.00479(4) & 0.00912(3) &
     {0.00513(4)} &   {0.00909(2)} \\
                   y     & 0.05003(4) & 0.03986(4) & 0.05634(5) & 0.04365(5) &
     {0.05536(4)} &   {0.04350(5)} \\
                   z     & 0.352152(13) & 0.351206(12) & 0.352293(17) & 0.350910(16) &
     {0.35217(2)} &   {0.350894(13)} \\
                   U$_{11}$ & 0.00420(6) & 0.00988(5) & 0.00485(7) & 0.01022(8) &
     {0.00599(7)} &   {0.01034(6)} \\
                   U$_{22}$ & 0.00526(8) & 0.01163(7) & 0.00530(9) & 0.01104(9) &
     {0.00678(8)} &   {0.01153(8)} \\
                   U$_{33}$ & 0.00261(5) & 0.00530(5) & 0.00360(7) & 0.00523(7) &
     {0.00460(8)} &   {0.00537(6)} \\
                   U$_{12}$ & -0.00092(5) & -0.00239(4) & -0.00077(7) & -0.00227(5) &
     {-0.00081(6)} &   {-0.00220(3)} \\
                   U$_{13}$ & -0.00002(4) & 0.00020(3) & 0.00007(6) & 0.00032(4) &
     {0.00004(7)} &   {0.00030(3)} \\
                   U$_{23}$ & -0.00047(4) & -0.00047(4) & -0.00022(6) & -0.00023(6) &
     {-0.00021(6)} &   {-0.00026(4)} \\
                   occupancy &   {1} &   {1} &
     {1} &   {1} & 1     & 1 \\
     \hline
                   O1    &       &       &       &       &       &  \\
                   x     & 0.19908(12) & 0.20016(9) & 0.19577(15) & 0.19758(12) &
     {0.19604(16)} &   {0.19770(9)} \\
                   y     & 0.29878(14) & 0.29890(10) & 0.30051(15) & 0.30081(12) &
     {0.30057(16)} &   {0.30061(9)} \\
                   z     & 0.02532(6) & 0.02228(5) & 0.02700(7) & 0.02325(6) &
     {0.02681(9)} &   {0.02322(4)} \\
                   U$_{11}$ & 0.0048(2) & 0.00732(15) & 0.0049(3) & 0.0072(2) &
     {0.0057(3)} &   {0.00734(16)} \\
                   U$_{22}$ & 0.0052(3) & 0.00788(18) & 0.0050(3) & 0.0068(2) &
     {0.0059(3)} &   {0.00695(16)} \\
                   U$_{33}$ & 0.0060(2) & 0.01175(18) & 0.0061(3) & 0.0108(2) &
     {0.0074(3)} &   {0.01106(17)} \\
                   U$_{12}$ & -0.0024(2) & -0.00339(13) & -0.0019(2) & -0.00312(18) &
     {-0.0016(2)} &   {-0.00263(12)} \\
                   U$_{13}$ & -0.00055(17) & 0.00025(14) & 0.0006(2) & 0.00088(18) &
     {0.0005(3)} &   {0.00107(13)} \\
                   U$_{23}$ & 0.00039(19) & -0.00013(15) & 0.0000(2) & -0.0010(2) &
     {-0.0004(3)} &   {-0.00078(14)} \\
                   occupancy & 1.033(4) & 1.028(3) & 1.016(4) & 1.014(4) &
     {1.012(5)} &   {1.009(4)} \\
     \hline
                   O2    &       &       &       &       &       &  \\
                   x     & -0.06456(13) & -0.05697(12) & -0.06762(18) & -0.05808(18) &
     {-0.06701(18)} &   {-0.05813(13)} \\
                   y     & -0.01939(14) & -0.01561(11) & -0.02114(16) & -0.01661(13) &
     {-0.02093(16)} &   {-0.01684(10)} \\
                   z     & 0.16450(6) & 0.16468(4) & 0.16458(7) & 0.16488(6) &
     {0.16453(9)} &   {0.16491(4)} \\
                   U$_{11}$ & 0.0064(2) & 0.0121(2) & 0.0071(3) & 0.0128(3) &
     {0.0082(3)} &   {0.0123(3)} \\
                   U$_{22}$ & 0.0072(3) & 0.0132(2) & 0.0067(3) & 0.0125(3) &
     {0.0080(3)} &   {0.0125(2)} \\
                   U$_{33}$ & 0.0034(2) & 0.00558(16) & 0.0045(3) & 0.0055(2) &
     {0.0059(3)} &   {0.00531(16)} \\
                   U$_{12}$ & 0.00075(20) & 0.00089(15) & 0.0005(2) & 0.0009(2) &
     {0.0004(2)} &   {0.00108(15)} \\
                   U$_{13}$ & 0.00013(16) & 0.00065(14) & 0.0004(2) & 0.00065(19) &
     {-0.0001(3)} &   {0.00066(15)} \\
                   U$_{23}$ & 0.00001(18) & 0.00023(13) & 0.0002(2) & 0.0002(2) &
     {0.0002(3)} &   {-0.00008(13)} \\
                   occupancy & 1.046(5) & 1.040(5) & 1.025(5) & 1.029(5) &
     {1.028(6)} &   {1.012(5)} \\ \hline  \hline

           \end{tabular}
           \label{tab:ref}
    \end{table*}

\begin{table}
\caption{Structural and physical characteristics of crystals containing different amounts of Ti.}
\vspace{5mm}
\begin{tabular}{l  c  c  c }
\hline
\hline
    & 0 \% Ti          & 1 \% Ti     & 10 \% Ti            \\ \hline
    $T_N$ (K)& $112.6(2)$  & $112.3(2)$   & $107.2(3)$   \\
    $T_{MI}$ (K)& $362(1)$  & $358(1)$   & $-$   \\ \hline
    \textbf{300 K} & & & \\
$a$ (\AA)& $5.4098(3)$  & $5.4098(3)$   & $5.4247(4)$   \\
$b$ (\AA)& $5.4691(4)$  & $5.4683(4)$   & $5.4585(5)$   \\
$c$ (\AA)& $11.9745(6)$  & $11.9781(9)$   & $11.9536(9)$   \\
$\epsilon$ & $0.00545(14)$  & $0.00537(17)$   & $0.00311(17)$  \\
$Ru-O1_{aver.}$ (\AA) & 1.9816(8) & 1.9852(7) & 1.9779(5)\\
$Ru-O2$ (\AA) & 2.0018(9) & 2.003(8) & 1.9932(5)\\
$O-O||a $ (\AA) & 2.815(1)	& 2.817(1)	& 2.815(1) \\
$O-O||b $ (\AA) 	& 2.792(1) & 2.792(1)	& 2.782(1) \\
$\theta-O1$ (deg)& 11.253(17) & 11.38(2) & 10.896(12) \\
$\theta-O2$ (deg)& 9.278(20) & 9.423(2) & 9.260(9) \\
$\phi$ (deg)& 11.628(16) & 11.666(13) & 11.171(11) \\\hline

    \textbf{100 K} & & & \\
$a$ (\AA)& $5.377(11)$  & $5.3957(3)$   & $5.4189(4)$   \\
$b$ (\AA)& $5.5915(12)$  & $5.6023(3)$   & $5.5483(5)$   \\
$c$ (\AA)& $11.789(3)$  & $11.7725(7)$   & $11.7982(10)$   \\
$\epsilon$ & $0.0176(2)$  & $0.0188(4)$   & $0.0118(6)$  \\
$Ru-O1_{aver.}$ (\AA) & 2.0099(6)  & 2.0120(6)	& 1.9996(5)	\\
$Ru-O2$ (\AA) & 1.9765(12)	& 1.9751(8)	& 1.9753(7) \\
$O-O||a $ (\AA) & 2.829(1)	& 2.829(1)	& 2.827(1) \\
$O-O||b $ (\AA) & 2.856(1)	& 2.862(1)	& 2.829(1) \\
$\theta-O1$ (deg) & 12.912(21) 	& 12.986(16)	& 12.201(14) \\
$\theta-O2$ (deg)& 11.089(19) & 11.199(19) & 10.680(11) \\
$\phi$ (deg)& 11.801(18)	& 11.823(17) & 11.287(15) \\\hline \hline
\end{tabular}
\label{tab:char}
\end{table}

\section{Magnetic structure determination of  Ca$_2$Ru$_{0.99}$Ti$_{0.01}$O$_4$}

\begin{figure}
 \includegraphics[width=0.7\columnwidth]{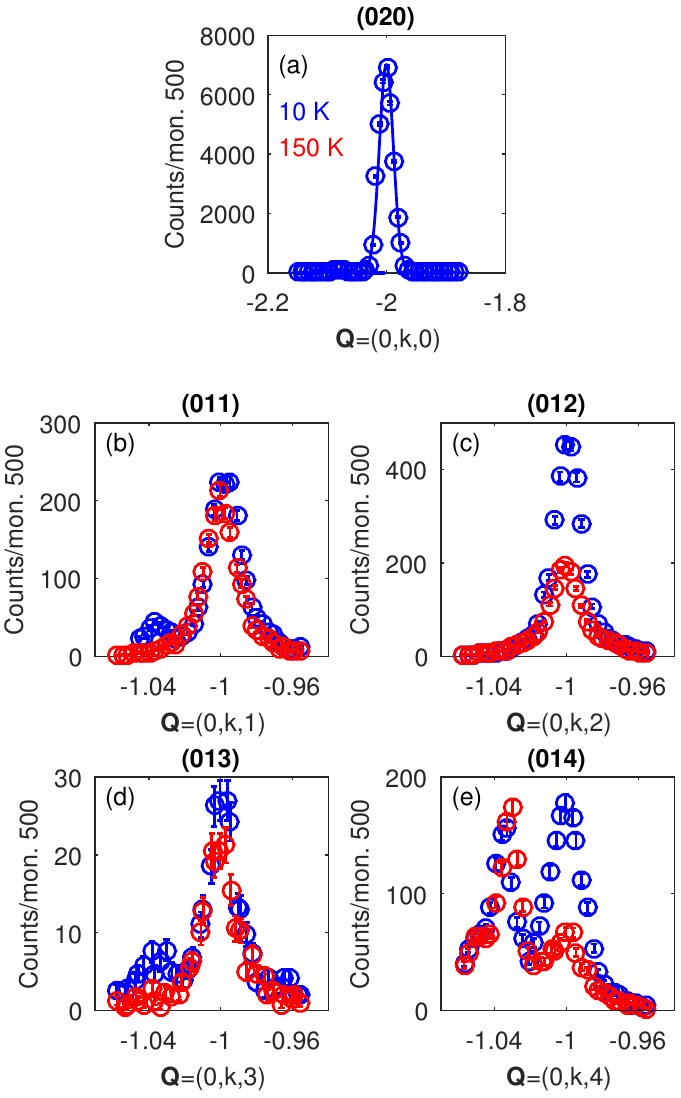}%
 \caption{\label{magn-str}(color online)   Elastic neutron scattering scans of 1TiB across (a) $\textbf Q=(0,2,0)$, (b) $\textbf Q=(0,1,0)$, (c) $\textbf Q=(0,1,l)$, (d) $\textbf Q=(0,1,2)$, (e) $\textbf Q=(0,1,2)$. Blue color denotes scans taken in the magnetic phase at 10~K and red color scans taken in the non magnetic phase at 150~K.}
 \end{figure}

 The crystal 1TiB used in the previous neutron scattering study \cite{Kunkemoeller2015} was further investigated on the thermal TAS IN3 at the ILL. The crystal was mounted in the [010]/[001] orientation into a ILL orange cryostat, $k_f$ was set to 2.662~\AA$^{-1}$ for all scans and a pyrolitic graphite filter was used to suppress higher order wave length. Rocking scans revealed the mosaic spread to be less than 0.5$^\circ$. With the good resolution of the IN3 it is easily possible to resolve the Bragg peaks from the two twins present in the crystal. The twinning ratio amounts to 20:1, which is deduced from rocking scans of strong nuclear Bragg reflections at the 2$\Theta$ scattering angles of the (020) and (200) reflections. Figure~\ref{magn-str}(a) shows a longitudinal scan over the (020) reflection, the smaller twin is barely seen in the tail. Scans across the positions, where magnetic scattering can be expected, are shown in (b-e). There several scans of the kind (01l) are compared at 10~K, in the antiferromagnetic phase, and at 150~K, in the non magnetic phase. It is clearly seen, that the minority twin gains intensity for an odd $l$ and not for an even $l$. The majority twin gains intensity for even l and much less, but also clearly detectable, for odd l. The (012) and (014) reflections are magnetic Bragg peaks for the $B$-centered phase but not for the $A$-centered one. So the strong gain of intensity of these reflections in the antiferromagnetic phase reveals an almost exclusive $B$-centered phase. In opposite the (011) and (013) peaks are Bragg peaks in the $A$-centered phase, but not in the $B$-centered one. So the gain of intensity  at low temperatures of those peaks points to the $A$-centered phase, which is however much weaker than the gain of the Bragg peak intensities of the $B$-centered phase. So the main stacking scheme of this crystal is the $B$-centered one.

\newpage


\begin{thebibliography}{53}%
\makeatletter
\providecommand \@ifxundefined [1]{%
 \@ifx{#1\undefined}
}%
\providecommand \@ifnum [1]{%
 \ifnum #1\expandafter \@firstoftwo
 \else \expandafter \@secondoftwo
 \fi
}%
\providecommand \@ifx [1]{%
 \ifx #1\expandafter \@firstoftwo
 \else \expandafter \@secondoftwo
 \fi
}%
\providecommand \natexlab [1]{#1}%
\providecommand \enquote  [1]{``#1''}%
\providecommand \bibnamefont  [1]{#1}%
\providecommand \bibfnamefont [1]{#1}%
\providecommand \citenamefont [1]{#1}%
\providecommand \href@noop [0]{\@secondoftwo}%
\providecommand \href [0]{\begingroup \@sanitize@url \@href}%
\providecommand \@href[1]{\@@startlink{#1}\@@href}%
\providecommand \@@href[1]{\endgroup#1\@@endlink}%
\providecommand \@sanitize@url [0]{\catcode `\\12\catcode `\$12\catcode
  `\&12\catcode `\#12\catcode `\^12\catcode `\_12\catcode `\%12\relax}%
\providecommand \@@startlink[1]{}%
\providecommand \@@endlink[0]{}%
\providecommand \url  [0]{\begingroup\@sanitize@url \@url }%
\providecommand \@url [1]{\endgroup\@href {#1}{\urlprefix }}%
\providecommand \urlprefix  [0]{URL }%
\providecommand \Eprint [0]{\href }%
\providecommand \doibase [0]{http://dx.doi.org/}%
\providecommand \selectlanguage [0]{\@gobble}%
\providecommand \bibinfo  [0]{\@secondoftwo}%
\providecommand \bibfield  [0]{\@secondoftwo}%
\providecommand \translation [1]{[#1]}%
\providecommand \BibitemOpen [0]{}%
\providecommand \bibitemStop [0]{}%
\providecommand \bibitemNoStop [0]{.\EOS\space}%
\providecommand \EOS [0]{\spacefactor3000\relax}%
\providecommand \BibitemShut  [1]{\csname bibitem#1\endcsname}%
\let\auto@bib@innerbib\@empty
\bibitem [{\citenamefont {Nakatsuji}\ \emph {et~al.}(1997)\citenamefont
  {Nakatsuji}, \citenamefont {Ikeda},\ and\ \citenamefont
  {Maeno}}]{Nakatsuji1997}%
  \BibitemOpen
  \bibfield  {author} {\bibinfo {author} {\bibfnamefont {S.}~\bibnamefont
  {Nakatsuji}}, \bibinfo {author} {\bibfnamefont {S.}~\bibnamefont {Ikeda}}, \
  and\ \bibinfo {author} {\bibfnamefont {Y.}~\bibnamefont {Maeno}},\ }\href
  {\doibase 10.1143/JPSJ.66.1868} {\bibfield  {journal} {\bibinfo  {journal}
  {J. Phys. Soc. Jpn.}\ }\textbf {\bibinfo {volume} {66}},\ \bibinfo {pages}
  {1868} (\bibinfo {year} {1997})}\BibitemShut {NoStop}%
\bibitem [{\citenamefont {Nakatsuji}\ and\ \citenamefont
  {Maeno}(2000)}]{Nakatsuji2000a}%
  \BibitemOpen
  \bibfield  {author} {\bibinfo {author} {\bibfnamefont {S.}~\bibnamefont
  {Nakatsuji}}\ and\ \bibinfo {author} {\bibfnamefont {Y.}~\bibnamefont
  {Maeno}},\ }\href {\doibase 10.1103/PhysRevLett.84.2666} {\bibfield
  {journal} {\bibinfo  {journal} {Phys. Rev. Lett.}\ }\textbf {\bibinfo
  {volume} {84}},\ \bibinfo {pages} {2666} (\bibinfo {year}
  {2000})}\BibitemShut {NoStop}%
\bibitem [{\citenamefont {Friedt}\ \emph {et~al.}(2001)\citenamefont {Friedt},
  \citenamefont {Braden}, \citenamefont {Andr\'e}, \citenamefont {Adelmann},
  \citenamefont {Nakatsuji},\ and\ \citenamefont {Maeno}}]{Friedt2001}%
  \BibitemOpen
  \bibfield  {author} {\bibinfo {author} {\bibfnamefont {O.}~\bibnamefont
  {Friedt}}, \bibinfo {author} {\bibfnamefont {M.}~\bibnamefont {Braden}},
  \bibinfo {author} {\bibfnamefont {G.}~\bibnamefont {Andr\'e}}, \bibinfo
  {author} {\bibfnamefont {P.}~\bibnamefont {Adelmann}}, \bibinfo {author}
  {\bibfnamefont {S.}~\bibnamefont {Nakatsuji}}, \ and\ \bibinfo {author}
  {\bibfnamefont {Y.}~\bibnamefont {Maeno}},\ }\href {\doibase
  10.1103/PhysRevB.63.174432} {\bibfield  {journal} {\bibinfo  {journal} {Phys.
  Rev. B}\ }\textbf {\bibinfo {volume} {63}},\ \bibinfo {pages} {174432}
  (\bibinfo {year} {2001})}\BibitemShut {NoStop}%
\bibitem [{\citenamefont {Carlo}\ \emph {et~al.}(2012)\citenamefont {Carlo},
  \citenamefont {Goko}, \citenamefont {Gat-Malureanu}, \citenamefont {Russo},
  \citenamefont {Savici}, \citenamefont {Aczel}, \citenamefont {MacDougall},
  \citenamefont {Rodriguez}, \citenamefont {Williams}, \citenamefont {Luke},
  \citenamefont {Wiebe}, \citenamefont {Yoshida}, \citenamefont {Nakatsuji},
  \citenamefont {Maeno}, \citenamefont {Taniguchi},\ and\ \citenamefont
  {Uemura}}]{Carlo2012}%
  \BibitemOpen
  \bibfield  {author} {\bibinfo {author} {\bibfnamefont {J.~P.}\ \bibnamefont
  {Carlo}}, \bibinfo {author} {\bibfnamefont {T.}~\bibnamefont {Goko}},
  \bibinfo {author} {\bibfnamefont {I.~M.}\ \bibnamefont {Gat-Malureanu}},
  \bibinfo {author} {\bibfnamefont {P.~L.}\ \bibnamefont {Russo}}, \bibinfo
  {author} {\bibfnamefont {A.~T.}\ \bibnamefont {Savici}}, \bibinfo {author}
  {\bibfnamefont {A.~A.}\ \bibnamefont {Aczel}}, \bibinfo {author}
  {\bibfnamefont {G.~J.}\ \bibnamefont {MacDougall}}, \bibinfo {author}
  {\bibfnamefont {J.~A.}\ \bibnamefont {Rodriguez}}, \bibinfo {author}
  {\bibfnamefont {T.~J.}\ \bibnamefont {Williams}}, \bibinfo {author}
  {\bibfnamefont {G.~M.}\ \bibnamefont {Luke}}, \bibinfo {author}
  {\bibfnamefont {C.~R.}\ \bibnamefont {Wiebe}}, \bibinfo {author}
  {\bibfnamefont {Y.}~\bibnamefont {Yoshida}}, \bibinfo {author} {\bibfnamefont
  {S.}~\bibnamefont {Nakatsuji}}, \bibinfo {author} {\bibfnamefont
  {Y.}~\bibnamefont {Maeno}}, \bibinfo {author} {\bibfnamefont
  {T.}~\bibnamefont {Taniguchi}}, \ and\ \bibinfo {author} {\bibfnamefont
  {Y.~J.}\ \bibnamefont {Uemura}},\ }\href@noop {} {\bibfield  {journal}
  {\bibinfo  {journal} {Nature Materials}\ }\textbf {\bibinfo {volume} {11}},\
  \bibinfo {pages} {323} (\bibinfo {year} {2012})}\BibitemShut {NoStop}%
\bibitem [{\citenamefont {Nakatsuji}\ \emph {et~al.}(2004)\citenamefont
  {Nakatsuji}, \citenamefont {Dobrosavljevi\ifmmode~\acute{c}\else \'{c}\fi{}},
  \citenamefont {Tanaskovi\ifmmode~\acute{c}\else \'{c}\fi{}}, \citenamefont
  {Minakata}, \citenamefont {Fukazawa},\ and\ \citenamefont
  {Maeno}}]{Nakatsuji2004}%
  \BibitemOpen
  \bibfield  {author} {\bibinfo {author} {\bibfnamefont {S.}~\bibnamefont
  {Nakatsuji}}, \bibinfo {author} {\bibfnamefont {V.}~\bibnamefont
  {Dobrosavljevi\ifmmode~\acute{c}\else \'{c}\fi{}}}, \bibinfo {author}
  {\bibfnamefont {D.}~\bibnamefont {Tanaskovi\ifmmode~\acute{c}\else
  \'{c}\fi{}}}, \bibinfo {author} {\bibfnamefont {M.}~\bibnamefont {Minakata}},
  \bibinfo {author} {\bibfnamefont {H.}~\bibnamefont {Fukazawa}}, \ and\
  \bibinfo {author} {\bibfnamefont {Y.}~\bibnamefont {Maeno}},\ }\href
  {\doibase 10.1103/PhysRevLett.93.146401} {\bibfield  {journal} {\bibinfo
  {journal} {Phys. Rev. Lett.}\ }\textbf {\bibinfo {volume} {93}},\ \bibinfo
  {pages} {146401} (\bibinfo {year} {2004})}\BibitemShut {NoStop}%
\bibitem [{\citenamefont {Maeno}\ \emph {et~al.}(1994)\citenamefont {Maeno},
  \citenamefont {Hashimoto}, \citenamefont {Yoshida}, \citenamefont
  {Nishizaki}, \citenamefont {Fujita}, \citenamefont {Bednorz},\ and\
  \citenamefont {Lichtenberg}}]{Y.Maeno1994}%
  \BibitemOpen
  \bibfield  {author} {\bibinfo {author} {\bibfnamefont {Y.}~\bibnamefont
  {Maeno}}, \bibinfo {author} {\bibfnamefont {H.}~\bibnamefont {Hashimoto}},
  \bibinfo {author} {\bibfnamefont {K.}~\bibnamefont {Yoshida}}, \bibinfo
  {author} {\bibfnamefont {S.}~\bibnamefont {Nishizaki}}, \bibinfo {author}
  {\bibfnamefont {T.}~\bibnamefont {Fujita}}, \bibinfo {author} {\bibfnamefont
  {J.~G.}\ \bibnamefont {Bednorz}}, \ and\ \bibinfo {author} {\bibfnamefont
  {F.}~\bibnamefont {Lichtenberg}},\ }\href@noop {} {\bibfield  {journal}
  {\bibinfo  {journal} {Nature (London)}\ }\textbf {\bibinfo {volume} {372}},\
  \bibinfo {pages} {532} (\bibinfo {year} {1994})}\BibitemShut {NoStop}%
\bibitem [{\citenamefont {Ishida}\ \emph {et~al.}(1998)\citenamefont {Ishida},
  \citenamefont {Mukuda}, \citenamefont {Kitaoka}, \citenamefont {Asayama},
  \citenamefont {Mao}, \citenamefont {Mori},\ and\ \citenamefont
  {Maeno}}]{Ishida1998}%
  \BibitemOpen
  \bibfield  {author} {\bibinfo {author} {\bibfnamefont {K.}~\bibnamefont
  {Ishida}}, \bibinfo {author} {\bibfnamefont {H.}~\bibnamefont {Mukuda}},
  \bibinfo {author} {\bibfnamefont {Y.}~\bibnamefont {Kitaoka}}, \bibinfo
  {author} {\bibfnamefont {K.}~\bibnamefont {Asayama}}, \bibinfo {author}
  {\bibfnamefont {Z.~Q.}\ \bibnamefont {Mao}}, \bibinfo {author} {\bibfnamefont
  {Y.}~\bibnamefont {Mori}}, \ and\ \bibinfo {author} {\bibfnamefont
  {Y.}~\bibnamefont {Maeno}},\ }\href@noop {} {\bibfield  {journal} {\bibinfo
  {journal} {Nature (London)}\ }\textbf {\bibinfo {volume} {396}},\ \bibinfo
  {pages} {658} (\bibinfo {year} {1998})}\BibitemShut {NoStop}%
\bibitem [{\citenamefont {Luke}\ \emph {et~al.}(1998)\citenamefont {Luke},
  \citenamefont {Fudamoto}, \citenamefont {Kojima}, \citenamefont {Larkin},
  \citenamefont {Merrin}, \citenamefont {Nachumi}, \citenamefont {Uemura},
  \citenamefont {Maeno}, \citenamefont {Mao}, \citenamefont {Mori},
  \citenamefont {Nakamura},\ and\ \citenamefont {Sigrist}}]{Luke1998}%
  \BibitemOpen
  \bibfield  {author} {\bibinfo {author} {\bibfnamefont {G.~M.}\ \bibnamefont
  {Luke}}, \bibinfo {author} {\bibfnamefont {Y.}~\bibnamefont {Fudamoto}},
  \bibinfo {author} {\bibfnamefont {K.~M.}\ \bibnamefont {Kojima}}, \bibinfo
  {author} {\bibfnamefont {M.~I.}\ \bibnamefont {Larkin}}, \bibinfo {author}
  {\bibfnamefont {J.}~\bibnamefont {Merrin}}, \bibinfo {author} {\bibfnamefont
  {B.}~\bibnamefont {Nachumi}}, \bibinfo {author} {\bibfnamefont {Y.~J.}\
  \bibnamefont {Uemura}}, \bibinfo {author} {\bibfnamefont {Y.}~\bibnamefont
  {Maeno}}, \bibinfo {author} {\bibfnamefont {Z.~Q.}\ \bibnamefont {Mao}},
  \bibinfo {author} {\bibfnamefont {Y.}~\bibnamefont {Mori}}, \bibinfo {author}
  {\bibfnamefont {H.}~\bibnamefont {Nakamura}}, \ and\ \bibinfo {author}
  {\bibfnamefont {M.}~\bibnamefont {Sigrist}},\ }\href@noop {} {\bibfield
  {journal} {\bibinfo  {journal} {Nature (London)}\ }\textbf {\bibinfo {volume}
  {394}},\ \bibinfo {pages} {558} (\bibinfo {year} {1998})}\BibitemShut
  {NoStop}%
\bibitem [{\citenamefont {Maeno}\ \emph {et~al.}(2012)\citenamefont {Maeno},
  \citenamefont {Kittaka}, \citenamefont {Nomura}, \citenamefont {Yonezawa},\
  and\ \citenamefont {Ishida}}]{Maeno2012}%
  \BibitemOpen
  \bibfield  {author} {\bibinfo {author} {\bibfnamefont {Y.}~\bibnamefont
  {Maeno}}, \bibinfo {author} {\bibfnamefont {S.}~\bibnamefont {Kittaka}},
  \bibinfo {author} {\bibfnamefont {T.}~\bibnamefont {Nomura}}, \bibinfo
  {author} {\bibfnamefont {S.}~\bibnamefont {Yonezawa}}, \ and\ \bibinfo
  {author} {\bibfnamefont {K.}~\bibnamefont {Ishida}},\ }\href@noop {}
  {\bibfield  {journal} {\bibinfo  {journal} {Journal of the Physical Society
  of Japan}\ }\textbf {\bibinfo {volume} {81}},\ \bibinfo {pages} {011009}
  (\bibinfo {year} {2012})}\BibitemShut {NoStop}%
\bibitem [{\citenamefont {Alexander}\ \emph {et~al.}(1999)\citenamefont
  {Alexander}, \citenamefont {Cao}, \citenamefont {Dobrosavljevic},
  \citenamefont {McCall}, \citenamefont {Crow}, \citenamefont {Lochner},\ and\
  \citenamefont {Guertin}}]{Alexander1999}%
  \BibitemOpen
  \bibfield  {author} {\bibinfo {author} {\bibfnamefont {C.~S.}\ \bibnamefont
  {Alexander}}, \bibinfo {author} {\bibfnamefont {G.}~\bibnamefont {Cao}},
  \bibinfo {author} {\bibfnamefont {V.}~\bibnamefont {Dobrosavljevic}},
  \bibinfo {author} {\bibfnamefont {S.}~\bibnamefont {McCall}}, \bibinfo
  {author} {\bibfnamefont {J.~E.}\ \bibnamefont {Crow}}, \bibinfo {author}
  {\bibfnamefont {E.}~\bibnamefont {Lochner}}, \ and\ \bibinfo {author}
  {\bibfnamefont {R.~P.}\ \bibnamefont {Guertin}},\ }\href {\doibase
  10.1103/PhysRevB.60.R8422} {\bibfield  {journal} {\bibinfo  {journal} {Phys.
  Rev. B}\ }\textbf {\bibinfo {volume} {60}},\ \bibinfo {pages} {R8422}
  (\bibinfo {year} {1999})}\BibitemShut {NoStop}%
\bibitem [{\citenamefont {Braden}\ \emph
  {et~al.}(1998{\natexlab{a}})\citenamefont {Braden}, \citenamefont {Andr\'e},
  \citenamefont {Nakatsuji},\ and\ \citenamefont {Maeno}}]{Braden1998}%
  \BibitemOpen
  \bibfield  {author} {\bibinfo {author} {\bibfnamefont {M.}~\bibnamefont
  {Braden}}, \bibinfo {author} {\bibfnamefont {G.}~\bibnamefont {Andr\'e}},
  \bibinfo {author} {\bibfnamefont {S.}~\bibnamefont {Nakatsuji}}, \ and\
  \bibinfo {author} {\bibfnamefont {Y.}~\bibnamefont {Maeno}},\ }\href
  {\doibase 10.1103/PhysRevB.58.847} {\bibfield  {journal} {\bibinfo  {journal}
  {Phys. Rev. B}\ }\textbf {\bibinfo {volume} {58}},\ \bibinfo {pages} {847}
  (\bibinfo {year} {1998}{\natexlab{a}})}\BibitemShut {NoStop}%
\bibitem [{\citenamefont {Mizokawa}\ \emph {et~al.}(2001)\citenamefont
  {Mizokawa}, \citenamefont {Tjeng}, \citenamefont {Sawatzky}, \citenamefont
  {Ghiringhelli}, \citenamefont {Tjernberg}, \citenamefont {Brookes},
  \citenamefont {Fukazawa}, \citenamefont {Nakatsuji},\ and\ \citenamefont
  {Maeno}}]{Mizokawa2001}%
  \BibitemOpen
  \bibfield  {author} {\bibinfo {author} {\bibfnamefont {T.}~\bibnamefont
  {Mizokawa}}, \bibinfo {author} {\bibfnamefont {L.~H.}\ \bibnamefont {Tjeng}},
  \bibinfo {author} {\bibfnamefont {G.~A.}\ \bibnamefont {Sawatzky}}, \bibinfo
  {author} {\bibfnamefont {G.}~\bibnamefont {Ghiringhelli}}, \bibinfo {author}
  {\bibfnamefont {O.}~\bibnamefont {Tjernberg}}, \bibinfo {author}
  {\bibfnamefont {N.~B.}\ \bibnamefont {Brookes}}, \bibinfo {author}
  {\bibfnamefont {H.}~\bibnamefont {Fukazawa}}, \bibinfo {author}
  {\bibfnamefont {S.}~\bibnamefont {Nakatsuji}}, \ and\ \bibinfo {author}
  {\bibfnamefont {Y.}~\bibnamefont {Maeno}},\ }\href {\doibase
  10.1103/PhysRevLett.87.077202} {\bibfield  {journal} {\bibinfo  {journal}
  {Phys. Rev. Lett.}\ }\textbf {\bibinfo {volume} {87}},\ \bibinfo {pages}
  {077202} (\bibinfo {year} {2001})}\BibitemShut {NoStop}%
\bibitem [{\citenamefont {Gorelov}\ \emph {et~al.}(2010)\citenamefont
  {Gorelov}, \citenamefont {Karolak}, \citenamefont {Wehling}, \citenamefont
  {Lechermann}, \citenamefont {Lichtenstein},\ and\ \citenamefont
  {Pavarini}}]{Gorelov2010}%
  \BibitemOpen
  \bibfield  {author} {\bibinfo {author} {\bibfnamefont {E.}~\bibnamefont
  {Gorelov}}, \bibinfo {author} {\bibfnamefont {M.}~\bibnamefont {Karolak}},
  \bibinfo {author} {\bibfnamefont {T.~O.}\ \bibnamefont {Wehling}}, \bibinfo
  {author} {\bibfnamefont {F.}~\bibnamefont {Lechermann}}, \bibinfo {author}
  {\bibfnamefont {A.~I.}\ \bibnamefont {Lichtenstein}}, \ and\ \bibinfo
  {author} {\bibfnamefont {E.}~\bibnamefont {Pavarini}},\ }\href {\doibase
  10.1103/PhysRevLett.104.226401} {\bibfield  {journal} {\bibinfo  {journal}
  {Phys. Rev. Lett.}\ }\textbf {\bibinfo {volume} {104}},\ \bibinfo {pages}
  {226401} (\bibinfo {year} {2010})}\BibitemShut {NoStop}%
\bibitem [{\citenamefont {{Sutter}}\ \emph {et~al.}(2016)\citenamefont
  {{Sutter}}, \citenamefont {{Fatuzzo}}, \citenamefont {{Moser}}, \citenamefont
  {{Kim}}, \citenamefont {{Fittipaldi}}, \citenamefont {{Vecchione}},
  \citenamefont {{Granata}}, \citenamefont {{Sassa}}, \citenamefont
  {{Cossalter}}, \citenamefont {{Gatti}}, \citenamefont {{Grioni}},
  \citenamefont {{R\o{}nnow}}, \citenamefont {{Plumb}}, \citenamefont {{Matt}},
  \citenamefont {{Shi}}, \citenamefont {{Hoesch}}, \citenamefont {{Kim}},
  \citenamefont {{Chang}}, \citenamefont {{Jeng}}, \citenamefont {{Jozwiak}},
  \citenamefont {{Bostwick}}, \citenamefont {{Rotenberg}}, \citenamefont
  {{Georges}}, \citenamefont {{Neupert}},\ and\ \citenamefont
  {{Chang}}}]{Sutter2016}%
  \BibitemOpen
  \bibfield  {author} {\bibinfo {author} {\bibfnamefont {D.}~\bibnamefont
  {{Sutter}}}, \bibinfo {author} {\bibfnamefont {C.~G.}\ \bibnamefont
  {{Fatuzzo}}}, \bibinfo {author} {\bibfnamefont {S.}~\bibnamefont {{Moser}}},
  \bibinfo {author} {\bibfnamefont {M.}~\bibnamefont {{Kim}}}, \bibinfo
  {author} {\bibfnamefont {R.}~\bibnamefont {{Fittipaldi}}}, \bibinfo {author}
  {\bibfnamefont {A.}~\bibnamefont {{Vecchione}}}, \bibinfo {author}
  {\bibfnamefont {V.}~\bibnamefont {{Granata}}}, \bibinfo {author}
  {\bibfnamefont {Y.}~\bibnamefont {{Sassa}}}, \bibinfo {author} {\bibfnamefont
  {F.}~\bibnamefont {{Cossalter}}}, \bibinfo {author} {\bibfnamefont
  {G.}~\bibnamefont {{Gatti}}}, \bibinfo {author} {\bibfnamefont
  {M.}~\bibnamefont {{Grioni}}}, \bibinfo {author} {\bibfnamefont {H.~M.}\
  \bibnamefont {{R\o{}nnow}}}, \bibinfo {author} {\bibfnamefont {N.~C.}\
  \bibnamefont {{Plumb}}}, \bibinfo {author} {\bibfnamefont {C.~E.}\
  \bibnamefont {{Matt}}}, \bibinfo {author} {\bibfnamefont {M.}~\bibnamefont
  {{Shi}}}, \bibinfo {author} {\bibfnamefont {M.}~\bibnamefont {{Hoesch}}},
  \bibinfo {author} {\bibfnamefont {T.~K.}\ \bibnamefont {{Kim}}}, \bibinfo
  {author} {\bibfnamefont {T.-R.}\ \bibnamefont {{Chang}}}, \bibinfo {author}
  {\bibfnamefont {H.-T.}\ \bibnamefont {{Jeng}}}, \bibinfo {author}
  {\bibfnamefont {C.}~\bibnamefont {{Jozwiak}}}, \bibinfo {author}
  {\bibfnamefont {A.}~\bibnamefont {{Bostwick}}}, \bibinfo {author}
  {\bibfnamefont {E.}~\bibnamefont {{Rotenberg}}}, \bibinfo {author}
  {\bibfnamefont {A.}~\bibnamefont {{Georges}}}, \bibinfo {author}
  {\bibfnamefont {T.}~\bibnamefont {{Neupert}}}, \ and\ \bibinfo {author}
  {\bibfnamefont {J.}~\bibnamefont {{Chang}}},\ }\href@noop {} {\bibfield
  {journal} {\bibinfo  {journal} {ArXiv e-prints}\ } (\bibinfo {year}
  {2016})},\ \Eprint {http://arxiv.org/abs/1610.02854} {arXiv:1610.02854
  [cond-mat.str-el]} \BibitemShut {NoStop}%
\bibitem [{\citenamefont {Jung}\ \emph {et~al.}(2003)\citenamefont {Jung},
  \citenamefont {Fang}, \citenamefont {He}, \citenamefont {Kaneko},
  \citenamefont {Okimoto},\ and\ \citenamefont {Tokura}}]{Jung2003}%
  \BibitemOpen
  \bibfield  {author} {\bibinfo {author} {\bibfnamefont {J.~H.}\ \bibnamefont
  {Jung}}, \bibinfo {author} {\bibfnamefont {Z.}~\bibnamefont {Fang}}, \bibinfo
  {author} {\bibfnamefont {J.~P.}\ \bibnamefont {He}}, \bibinfo {author}
  {\bibfnamefont {Y.}~\bibnamefont {Kaneko}}, \bibinfo {author} {\bibfnamefont
  {Y.}~\bibnamefont {Okimoto}}, \ and\ \bibinfo {author} {\bibfnamefont
  {Y.}~\bibnamefont {Tokura}},\ }\href {\doibase 10.1103/PhysRevLett.91.056403}
  {\bibfield  {journal} {\bibinfo  {journal} {Phys. Rev. Lett.}\ }\textbf
  {\bibinfo {volume} {91}},\ \bibinfo {pages} {056403} (\bibinfo {year}
  {2003})}\BibitemShut {NoStop}%
\bibitem [{\citenamefont {Hotta}\ and\ \citenamefont
  {Dagotto}(2001)}]{Hotta2001}%
  \BibitemOpen
  \bibfield  {author} {\bibinfo {author} {\bibfnamefont {T.}~\bibnamefont
  {Hotta}}\ and\ \bibinfo {author} {\bibfnamefont {E.}~\bibnamefont
  {Dagotto}},\ }\href {\doibase 10.1103/PhysRevLett.88.017201} {\bibfield
  {journal} {\bibinfo  {journal} {Phys. Rev. Lett.}\ }\textbf {\bibinfo
  {volume} {88}},\ \bibinfo {pages} {017201} (\bibinfo {year}
  {2001})}\BibitemShut {NoStop}%
\bibitem [{\citenamefont {Puchkov}\ \emph {et~al.}(1998)\citenamefont
  {Puchkov}, \citenamefont {Schabel}, \citenamefont {Basov}, \citenamefont
  {Startseva}, \citenamefont {Cao}, \citenamefont {Timusk},\ and\ \citenamefont
  {Shen}}]{Puchkov1998}%
  \BibitemOpen
  \bibfield  {author} {\bibinfo {author} {\bibfnamefont {A.~V.}\ \bibnamefont
  {Puchkov}}, \bibinfo {author} {\bibfnamefont {M.~C.}\ \bibnamefont
  {Schabel}}, \bibinfo {author} {\bibfnamefont {D.~N.}\ \bibnamefont {Basov}},
  \bibinfo {author} {\bibfnamefont {T.}~\bibnamefont {Startseva}}, \bibinfo
  {author} {\bibfnamefont {G.}~\bibnamefont {Cao}}, \bibinfo {author}
  {\bibfnamefont {T.}~\bibnamefont {Timusk}}, \ and\ \bibinfo {author}
  {\bibfnamefont {Z.-X.}\ \bibnamefont {Shen}},\ }\href {\doibase
  10.1103/PhysRevLett.81.2747} {\bibfield  {journal} {\bibinfo  {journal}
  {Phys. Rev. Lett.}\ }\textbf {\bibinfo {volume} {81}},\ \bibinfo {pages}
  {2747} (\bibinfo {year} {1998})}\BibitemShut {NoStop}%
\bibitem [{\citenamefont {Liebsch}\ and\ \citenamefont
  {Ishida}(2007)}]{Liebsch2007}%
  \BibitemOpen
  \bibfield  {author} {\bibinfo {author} {\bibfnamefont {A.}~\bibnamefont
  {Liebsch}}\ and\ \bibinfo {author} {\bibfnamefont {H.}~\bibnamefont
  {Ishida}},\ }\href {\doibase 10.1103/PhysRevLett.98.216403} {\bibfield
  {journal} {\bibinfo  {journal} {Phys. Rev. Lett.}\ }\textbf {\bibinfo
  {volume} {98}},\ \bibinfo {pages} {216403} (\bibinfo {year}
  {2007})}\BibitemShut {NoStop}%
\bibitem [{\citenamefont {Liebsch}(2003)}]{Liebsch2003}%
  \BibitemOpen
  \bibfield  {author} {\bibinfo {author} {\bibfnamefont {A.}~\bibnamefont
  {Liebsch}},\ }\href {http://stacks.iop.org/0295-5075/63/i=1/a=097} {\bibfield
   {journal} {\bibinfo  {journal} {EPL (Europhysics Letters)}\ }\textbf
  {\bibinfo {volume} {63}},\ \bibinfo {pages} {97} (\bibinfo {year}
  {2003})}\BibitemShut {NoStop}%
\bibitem [{\citenamefont {Anisimov}\ \emph {et~al.}(2002)\citenamefont
  {Anisimov}, \citenamefont {Nekrasov}, \citenamefont {Kondakov}, \citenamefont
  {Rice},\ and\ \citenamefont {Sigrist}}]{Anisimov2002}%
  \BibitemOpen
  \bibfield  {author} {\bibinfo {author} {\bibfnamefont {V.~I.}\ \bibnamefont
  {Anisimov}}, \bibinfo {author} {\bibfnamefont {I.~A.}\ \bibnamefont
  {Nekrasov}}, \bibinfo {author} {\bibfnamefont {D.~E.}\ \bibnamefont
  {Kondakov}}, \bibinfo {author} {\bibfnamefont {T.~M.}\ \bibnamefont {Rice}},
  \ and\ \bibinfo {author} {\bibfnamefont {M.}~\bibnamefont {Sigrist}},\ }\href
  {\doibase 10.1140/epjb/e20020021} {\bibfield  {journal} {\bibinfo  {journal}
  {The European Physical Journal B - Condensed Matter and Complex Systems}\
  }\textbf {\bibinfo {volume} {25}},\ \bibinfo {pages} {191} (\bibinfo {year}
  {2002})}\BibitemShut {NoStop}%
\bibitem [{\citenamefont {Khomskii}(2014)}]{Khomskii2014}%
  \BibitemOpen
  \bibfield  {author} {\bibinfo {author} {\bibfnamefont {D.~I.}\ \bibnamefont
  {Khomskii}},\ }\href@noop {} {\emph {\bibinfo {title} {Transition Metal
  Compounds}}}\ (\bibinfo  {publisher} {Cambridge University Press},\ \bibinfo
  {address} {Cambridge, England},\ \bibinfo {year} {2014})\BibitemShut
  {NoStop}%
\bibitem [{\citenamefont {Khaliullin}(2013)}]{Khaliullin2013}%
  \BibitemOpen
  \bibfield  {author} {\bibinfo {author} {\bibfnamefont {G.}~\bibnamefont
  {Khaliullin}},\ }\href {\doibase 10.1103/PhysRevLett.111.197201} {\bibfield
  {journal} {\bibinfo  {journal} {Phys. Rev. Lett.}\ }\textbf {\bibinfo
  {volume} {111}},\ \bibinfo {pages} {197201} (\bibinfo {year}
  {2013})}\BibitemShut {NoStop}%
\bibitem [{\citenamefont {Akbari}\ and\ \citenamefont
  {Khaliullin}(2014)}]{Akbari2014}%
  \BibitemOpen
  \bibfield  {author} {\bibinfo {author} {\bibfnamefont {A.}~\bibnamefont
  {Akbari}}\ and\ \bibinfo {author} {\bibfnamefont {G.}~\bibnamefont
  {Khaliullin}},\ }\href {\doibase 10.1103/PhysRevB.90.035137} {\bibfield
  {journal} {\bibinfo  {journal} {Phys. Rev. B}\ }\textbf {\bibinfo {volume}
  {90}},\ \bibinfo {pages} {035137} (\bibinfo {year} {2014})}\BibitemShut
  {NoStop}%
\bibitem [{\citenamefont {Kunkem\"oller}\ \emph {et~al.}(2015)\citenamefont
  {Kunkem\"oller}, \citenamefont {Khomskii}, \citenamefont {Steffens},
  \citenamefont {Piovano}, \citenamefont {Nugroho},\ and\ \citenamefont
  {Braden}}]{Kunkemoeller2015}%
  \BibitemOpen
  \bibfield  {author} {\bibinfo {author} {\bibfnamefont {S.}~\bibnamefont
  {Kunkem\"oller}}, \bibinfo {author} {\bibfnamefont {D.}~\bibnamefont
  {Khomskii}}, \bibinfo {author} {\bibfnamefont {P.}~\bibnamefont {Steffens}},
  \bibinfo {author} {\bibfnamefont {A.}~\bibnamefont {Piovano}}, \bibinfo
  {author} {\bibfnamefont {A.~A.}\ \bibnamefont {Nugroho}}, \ and\ \bibinfo
  {author} {\bibfnamefont {M.}~\bibnamefont {Braden}},\ }\href {\doibase
  10.1103/PhysRevLett.115.247201} {\bibfield  {journal} {\bibinfo  {journal}
  {Phys. Rev. Lett.}\ }\textbf {\bibinfo {volume} {115}},\ \bibinfo {pages}
  {247201} (\bibinfo {year} {2015})}\BibitemShut {NoStop}%
\bibitem [{\citenamefont {{Jain}}\ \emph {et~al.}(2015)\citenamefont {{Jain}},
  \citenamefont {{Krautloher}}, \citenamefont {{Porras}}, \citenamefont
  {{Ryu}}, \citenamefont {{Chen}}, \citenamefont {{Abernathy}}, \citenamefont
  {{Park}}, \citenamefont {{Ivanov}}, \citenamefont {{Chaloupka}},
  \citenamefont {{Khaliullin}}, \citenamefont {{Keimer}},\ and\ \citenamefont
  {{Kim}}}]{Jain}%
  \BibitemOpen
  \bibfield  {author} {\bibinfo {author} {\bibfnamefont {A.}~\bibnamefont
  {{Jain}}}, \bibinfo {author} {\bibfnamefont {M.}~\bibnamefont
  {{Krautloher}}}, \bibinfo {author} {\bibfnamefont {J.}~\bibnamefont
  {{Porras}}}, \bibinfo {author} {\bibfnamefont {G.~H.}\ \bibnamefont {{Ryu}}},
  \bibinfo {author} {\bibfnamefont {D.~P.}\ \bibnamefont {{Chen}}}, \bibinfo
  {author} {\bibfnamefont {D.~L.}\ \bibnamefont {{Abernathy}}}, \bibinfo
  {author} {\bibfnamefont {J.~T.}\ \bibnamefont {{Park}}}, \bibinfo {author}
  {\bibfnamefont {A.}~\bibnamefont {{Ivanov}}}, \bibinfo {author}
  {\bibfnamefont {J.}~\bibnamefont {{Chaloupka}}}, \bibinfo {author}
  {\bibfnamefont {G.}~\bibnamefont {{Khaliullin}}}, \bibinfo {author}
  {\bibfnamefont {B.}~\bibnamefont {{Keimer}}}, \ and\ \bibinfo {author}
  {\bibfnamefont {B.~J.}\ \bibnamefont {{Kim}}},\ }\href@noop {} {\bibfield
  {journal} {\bibinfo  {journal} {ArXiv e-prints}\ } (\bibinfo {year}
  {2015})},\ \Eprint {http://arxiv.org/abs/1510.07011} {arXiv:1510.07011
  [cond-mat.str-el]} \BibitemShut {NoStop}%
\bibitem [{\citenamefont {Fukazawa}\ \emph {et~al.}(2000)\citenamefont
  {Fukazawa}, \citenamefont {Nakatsuji},\ and\ \citenamefont
  {Maeno}}]{Fukazawa2000}%
  \BibitemOpen
  \bibfield  {author} {\bibinfo {author} {\bibfnamefont {H.}~\bibnamefont
  {Fukazawa}}, \bibinfo {author} {\bibfnamefont {S.}~\bibnamefont {Nakatsuji}},
  \ and\ \bibinfo {author} {\bibfnamefont {Y.}~\bibnamefont {Maeno}},\ }\href
  {\doibase http://dx.doi.org/10.1016/S0921-4526(99)00989-8} {\bibfield
  {journal} {\bibinfo  {journal} {Physica B: Condensed Matter}\ }\textbf
  {\bibinfo {volume} {281}},\ \bibinfo {pages} {613 } (\bibinfo {year}
  {2000})}\BibitemShut {NoStop}%
\bibitem [{\citenamefont {Kunkem\"oller}\ \emph {et~al.}(2016)\citenamefont
  {Kunkem\"oller}, \citenamefont {Sauer}, \citenamefont {Nugroho},\ and\
  \citenamefont {Braden}}]{Kunkemoeller2016}%
  \BibitemOpen
  \bibfield  {author} {\bibinfo {author} {\bibfnamefont {S.}~\bibnamefont
  {Kunkem\"oller}}, \bibinfo {author} {\bibfnamefont {F.}~\bibnamefont
  {Sauer}}, \bibinfo {author} {\bibfnamefont {A.~A.}\ \bibnamefont {Nugroho}},
  \ and\ \bibinfo {author} {\bibfnamefont {M.}~\bibnamefont {Braden}},\ }\href
  {\doibase 10.1002/crat.201600020} {\bibfield  {journal} {\bibinfo  {journal}
  {Crystal Research and Technology}\ }\textbf {\bibinfo {volume} {51}},\
  \bibinfo {pages} {299} (\bibinfo {year} {2016})}\BibitemShut {NoStop}%
\bibitem [{\citenamefont
  {Rodr\'{i}guez-Carvajal}(1993)}]{Rodriguez-Carvajal1993}%
  \BibitemOpen
  \bibfield  {author} {\bibinfo {author} {\bibfnamefont {J.}~\bibnamefont
  {Rodr\'{i}guez-Carvajal}},\ }\href {\doibase
  http://dx.doi.org/10.1016/0921-4526(93)90108-I} {\bibfield  {journal}
  {\bibinfo  {journal} {Physica B: Condensed Matter}\ }\textbf {\bibinfo
  {volume} {192}},\ \bibinfo {pages} {55 } (\bibinfo {year}
  {1993})}\BibitemShut {NoStop}%
\bibitem [{\citenamefont {Toth}\ and\ \citenamefont {Lake}(2015)}]{toth2015}%
  \BibitemOpen
  \bibfield  {author} {\bibinfo {author} {\bibfnamefont {S.}~\bibnamefont
  {Toth}}\ and\ \bibinfo {author} {\bibfnamefont {B.}~\bibnamefont {Lake}},\
  }\href {http://stacks.iop.org/0953-8984/27/i=16/a=166002} {\bibfield
  {journal} {\bibinfo  {journal} {Journal of Physics: Condensed Matter}\
  }\textbf {\bibinfo {volume} {27}},\ \bibinfo {pages} {166002} (\bibinfo
  {year} {2015})}\BibitemShut {NoStop}%
\bibitem [{\citenamefont {Chatterji}(2005)}]{Chatterji2005}%
  \BibitemOpen
  \bibinfo {editor} {\bibfnamefont {T.}~\bibnamefont {Chatterji}},\ ed.,\
  \href@noop {} {\emph {\bibinfo {title} {Neutron Scattering from Magnetic
  Materials}}}\ (\bibinfo  {publisher} {Elsevier Science},\ \bibinfo {year}
  {2005})\BibitemShut {NoStop}%
\bibitem [{\citenamefont {Qureshi}\ \emph {et~al.}(2012)\citenamefont
  {Qureshi}, \citenamefont {Steffens}, \citenamefont {Wurmehl}, \citenamefont
  {Aswartham}, \citenamefont {B\"uchner},\ and\ \citenamefont
  {Braden}}]{Qureshi2012}%
  \BibitemOpen
  \bibfield  {author} {\bibinfo {author} {\bibfnamefont {N.}~\bibnamefont
  {Qureshi}}, \bibinfo {author} {\bibfnamefont {P.}~\bibnamefont {Steffens}},
  \bibinfo {author} {\bibfnamefont {S.}~\bibnamefont {Wurmehl}}, \bibinfo
  {author} {\bibfnamefont {S.}~\bibnamefont {Aswartham}}, \bibinfo {author}
  {\bibfnamefont {B.}~\bibnamefont {B\"uchner}}, \ and\ \bibinfo {author}
  {\bibfnamefont {M.}~\bibnamefont {Braden}},\ }\href {\doibase
  10.1103/PhysRevB.86.060410} {\bibfield  {journal} {\bibinfo  {journal} {Phys.
  Rev. B}\ }\textbf {\bibinfo {volume} {86}},\ \bibinfo {pages} {060410}
  (\bibinfo {year} {2012})}\BibitemShut {NoStop}%
\bibitem [{\citenamefont {Randall}\ and\ \citenamefont
  {Ward}(1959)}]{Randall1959}%
  \BibitemOpen
  \bibfield  {author} {\bibinfo {author} {\bibfnamefont {J.~J.}\ \bibnamefont
  {Randall}}\ and\ \bibinfo {author} {\bibfnamefont {R.}~\bibnamefont {Ward}},\
  }\href@noop {} {\bibfield  {journal} {\bibinfo  {journal} {J.Am.Chem.Soc}\
  }\textbf {\bibinfo {volume} {81}},\ \bibinfo {pages} {2629} (\bibinfo {year}
  {1959})}\BibitemShut {NoStop}%
\bibitem [{\citenamefont {Braden}\ \emph
  {et~al.}(1998{\natexlab{b}})\citenamefont {Braden}, \citenamefont
  {Reichardt}, \citenamefont {Nishizaki}, \citenamefont {Mori},\ and\
  \citenamefont {Maeno}}]{Braden1998a}%
  \BibitemOpen
  \bibfield  {author} {\bibinfo {author} {\bibfnamefont {M.}~\bibnamefont
  {Braden}}, \bibinfo {author} {\bibfnamefont {W.}~\bibnamefont {Reichardt}},
  \bibinfo {author} {\bibfnamefont {S.}~\bibnamefont {Nishizaki}}, \bibinfo
  {author} {\bibfnamefont {Y.}~\bibnamefont {Mori}}, \ and\ \bibinfo {author}
  {\bibfnamefont {Y.}~\bibnamefont {Maeno}},\ }\href {\doibase
  10.1103/PhysRevB.57.1236} {\bibfield  {journal} {\bibinfo  {journal} {Phys.
  Rev. B}\ }\textbf {\bibinfo {volume} {57}},\ \bibinfo {pages} {1236}
  (\bibinfo {year} {1998}{\natexlab{b}})}\BibitemShut {NoStop}%
\bibitem [{\citenamefont {Braden}\ \emph {et~al.}(2007)\citenamefont {Braden},
  \citenamefont {Reichardt}, \citenamefont {Sidis}, \citenamefont {Mao},\ and\
  \citenamefont {Maeno}}]{Braden2007}%
  \BibitemOpen
  \bibfield  {author} {\bibinfo {author} {\bibfnamefont {M.}~\bibnamefont
  {Braden}}, \bibinfo {author} {\bibfnamefont {W.}~\bibnamefont {Reichardt}},
  \bibinfo {author} {\bibfnamefont {Y.}~\bibnamefont {Sidis}}, \bibinfo
  {author} {\bibfnamefont {Z.}~\bibnamefont {Mao}}, \ and\ \bibinfo {author}
  {\bibfnamefont {Y.}~\bibnamefont {Maeno}},\ }\href {\doibase
  10.1103/PhysRevB.76.014505} {\bibfield  {journal} {\bibinfo  {journal} {Phys.
  Rev. B}\ }\textbf {\bibinfo {volume} {76}},\ \bibinfo {pages} {014505}
  (\bibinfo {year} {2007})}\BibitemShut {NoStop}%
\bibitem [{\citenamefont {Braden}\ \emph {et~al.}(1997)\citenamefont {Braden},
  \citenamefont {Moudden}, \citenamefont {Nishizaki}, \citenamefont {Maeno},\
  and\ \citenamefont {Fujita}}]{Braden1997}%
  \BibitemOpen
  \bibfield  {author} {\bibinfo {author} {\bibfnamefont {M.}~\bibnamefont
  {Braden}}, \bibinfo {author} {\bibfnamefont {A.~H.}\ \bibnamefont {Moudden}},
  \bibinfo {author} {\bibfnamefont {S.}~\bibnamefont {Nishizaki}}, \bibinfo
  {author} {\bibfnamefont {Y.}~\bibnamefont {Maeno}}, \ and\ \bibinfo {author}
  {\bibfnamefont {T.}~\bibnamefont {Fujita}},\ }\href {\doibase
  http://dx.doi.org/10.1016/S0921-4534(96)00637-5} {\bibfield  {journal}
  {\bibinfo  {journal} {Physica C: Superconductivity}\ }\textbf {\bibinfo
  {volume} {273}},\ \bibinfo {pages} {248 } (\bibinfo {year}
  {1997})}\BibitemShut {NoStop}%
\bibitem [{\citenamefont {Ke}\ \emph {et~al.}(2011)\citenamefont {Ke},
  \citenamefont {Peng}, \citenamefont {Singh}, \citenamefont {Hong},
  \citenamefont {Tian}, \citenamefont {Dela~Cruz},\ and\ \citenamefont
  {Mao}}]{Ke2011}%
  \BibitemOpen
  \bibfield  {author} {\bibinfo {author} {\bibfnamefont {X.}~\bibnamefont
  {Ke}}, \bibinfo {author} {\bibfnamefont {J.}~\bibnamefont {Peng}}, \bibinfo
  {author} {\bibfnamefont {D.~J.}\ \bibnamefont {Singh}}, \bibinfo {author}
  {\bibfnamefont {T.}~\bibnamefont {Hong}}, \bibinfo {author} {\bibfnamefont
  {W.}~\bibnamefont {Tian}}, \bibinfo {author} {\bibfnamefont {C.~R.}\
  \bibnamefont {Dela~Cruz}}, \ and\ \bibinfo {author} {\bibfnamefont {Z.~Q.}\
  \bibnamefont {Mao}},\ }\href {\doibase 10.1103/PhysRevB.84.201102} {\bibfield
   {journal} {\bibinfo  {journal} {Phys. Rev. B}\ }\textbf {\bibinfo {volume}
  {84}},\ \bibinfo {pages} {201102} (\bibinfo {year} {2011})}\BibitemShut
  {NoStop}%
\bibitem [{\citenamefont {Braden}\ \emph {et~al.}(2002)\citenamefont {Braden},
  \citenamefont {Friedt}, \citenamefont {Sidis}, \citenamefont {Bourges},
  \citenamefont {Minakata},\ and\ \citenamefont {Maeno}}]{Braden2002}%
  \BibitemOpen
  \bibfield  {author} {\bibinfo {author} {\bibfnamefont {M.}~\bibnamefont
  {Braden}}, \bibinfo {author} {\bibfnamefont {O.}~\bibnamefont {Friedt}},
  \bibinfo {author} {\bibfnamefont {Y.}~\bibnamefont {Sidis}}, \bibinfo
  {author} {\bibfnamefont {P.}~\bibnamefont {Bourges}}, \bibinfo {author}
  {\bibfnamefont {M.}~\bibnamefont {Minakata}}, \ and\ \bibinfo {author}
  {\bibfnamefont {Y.}~\bibnamefont {Maeno}},\ }\href {\doibase
  10.1103/PhysRevLett.88.197002} {\bibfield  {journal} {\bibinfo  {journal}
  {Phys. Rev. Lett.}\ }\textbf {\bibinfo {volume} {88}},\ \bibinfo {pages}
  {197002} (\bibinfo {year} {2002})}\BibitemShut {NoStop}%
\bibitem [{\citenamefont {Steffens}\ \emph {et~al.}(2009)\citenamefont
  {Steffens}, \citenamefont {Farrell}, \citenamefont {Price}, \citenamefont
  {Mackenzie}, \citenamefont {Sidis}, \citenamefont {Schmalzl},\ and\
  \citenamefont {Braden}}]{Steffens2009}%
  \BibitemOpen
  \bibfield  {author} {\bibinfo {author} {\bibfnamefont {P.}~\bibnamefont
  {Steffens}}, \bibinfo {author} {\bibfnamefont {J.}~\bibnamefont {Farrell}},
  \bibinfo {author} {\bibfnamefont {S.}~\bibnamefont {Price}}, \bibinfo
  {author} {\bibfnamefont {A.~P.}\ \bibnamefont {Mackenzie}}, \bibinfo {author}
  {\bibfnamefont {Y.}~\bibnamefont {Sidis}}, \bibinfo {author} {\bibfnamefont
  {K.}~\bibnamefont {Schmalzl}}, \ and\ \bibinfo {author} {\bibfnamefont
  {M.}~\bibnamefont {Braden}},\ }\href {\doibase 10.1103/PhysRevB.79.054422}
  {\bibfield  {journal} {\bibinfo  {journal} {Phys. Rev. B}\ }\textbf {\bibinfo
  {volume} {79}},\ \bibinfo {pages} {054422} (\bibinfo {year}
  {2009})}\BibitemShut {NoStop}%
\bibitem [{\citenamefont {Steffens}\ \emph {et~al.}(2005)\citenamefont
  {Steffens}, \citenamefont {Friedt}, \citenamefont {Alireza}, \citenamefont
  {Marshall}, \citenamefont {Schmidt}, \citenamefont {Nakamura}, \citenamefont
  {Nakatsuji}, \citenamefont {Maeno}, \citenamefont {Lengsdorf}, \citenamefont
  {Abd-Elmeguid},\ and\ \citenamefont {Braden}}]{Steffens2005}%
  \BibitemOpen
  \bibfield  {author} {\bibinfo {author} {\bibfnamefont {P.}~\bibnamefont
  {Steffens}}, \bibinfo {author} {\bibfnamefont {O.}~\bibnamefont {Friedt}},
  \bibinfo {author} {\bibfnamefont {P.}~\bibnamefont {Alireza}}, \bibinfo
  {author} {\bibfnamefont {W.~G.}\ \bibnamefont {Marshall}}, \bibinfo {author}
  {\bibfnamefont {W.}~\bibnamefont {Schmidt}}, \bibinfo {author} {\bibfnamefont
  {F.}~\bibnamefont {Nakamura}}, \bibinfo {author} {\bibfnamefont
  {S.}~\bibnamefont {Nakatsuji}}, \bibinfo {author} {\bibfnamefont
  {Y.}~\bibnamefont {Maeno}}, \bibinfo {author} {\bibfnamefont
  {R.}~\bibnamefont {Lengsdorf}}, \bibinfo {author} {\bibfnamefont {M.~M.}\
  \bibnamefont {Abd-Elmeguid}}, \ and\ \bibinfo {author} {\bibfnamefont
  {M.}~\bibnamefont {Braden}},\ }\href {\doibase 10.1103/PhysRevB.72.094104}
  {\bibfield  {journal} {\bibinfo  {journal} {Phys. Rev. B}\ }\textbf {\bibinfo
  {volume} {72}},\ \bibinfo {pages} {094104} (\bibinfo {year}
  {2005})}\BibitemShut {NoStop}%
\bibitem [{\citenamefont {Heilmann}\ \emph {et~al.}(1981)\citenamefont
  {Heilmann}, \citenamefont {Kjems}, \citenamefont {Endoh}, \citenamefont
  {Reiter}, \citenamefont {Shirane},\ and\ \citenamefont
  {Birgeneau}}]{Heilmann1981}%
  \BibitemOpen
  \bibfield  {author} {\bibinfo {author} {\bibfnamefont {I.~U.}\ \bibnamefont
  {Heilmann}}, \bibinfo {author} {\bibfnamefont {J.~K.}\ \bibnamefont {Kjems}},
  \bibinfo {author} {\bibfnamefont {Y.}~\bibnamefont {Endoh}}, \bibinfo
  {author} {\bibfnamefont {G.~F.}\ \bibnamefont {Reiter}}, \bibinfo {author}
  {\bibfnamefont {G.}~\bibnamefont {Shirane}}, \ and\ \bibinfo {author}
  {\bibfnamefont {R.~J.}\ \bibnamefont {Birgeneau}},\ }\href {\doibase
  10.1103/PhysRevB.24.3939} {\bibfield  {journal} {\bibinfo  {journal} {Phys.
  Rev. B}\ }\textbf {\bibinfo {volume} {24}},\ \bibinfo {pages} {3939}
  (\bibinfo {year} {1981})}\BibitemShut {NoStop}%
\bibitem [{\citenamefont {Wang}\ \emph {et~al.}(2013)\citenamefont {Wang},
  \citenamefont {Zhang}, \citenamefont {Wang}, \citenamefont {Luo},
  \citenamefont {Regnault}, \citenamefont {Dai},\ and\ \citenamefont
  {Li}}]{Wang2013}%
  \BibitemOpen
  \bibfield  {author} {\bibinfo {author} {\bibfnamefont {C.}~\bibnamefont
  {Wang}}, \bibinfo {author} {\bibfnamefont {R.}~\bibnamefont {Zhang}},
  \bibinfo {author} {\bibfnamefont {F.}~\bibnamefont {Wang}}, \bibinfo {author}
  {\bibfnamefont {H.}~\bibnamefont {Luo}}, \bibinfo {author} {\bibfnamefont
  {L.~P.}\ \bibnamefont {Regnault}}, \bibinfo {author} {\bibfnamefont
  {P.}~\bibnamefont {Dai}}, \ and\ \bibinfo {author} {\bibfnamefont
  {Y.}~\bibnamefont {Li}},\ }\href {\doibase 10.1103/PhysRevX.3.041036}
  {\bibfield  {journal} {\bibinfo  {journal} {Phys. Rev. X}\ }\textbf {\bibinfo
  {volume} {3}},\ \bibinfo {pages} {041036} (\bibinfo {year}
  {2013})}\BibitemShut {NoStop}%
\bibitem [{\citenamefont {Fidrysiak}(2016)}]{Fidrysiak2016}%
  \BibitemOpen
  \bibfield  {author} {\bibinfo {author} {\bibfnamefont {M.}~\bibnamefont
  {Fidrysiak}},\ }\href {\doibase 10.1140/epjb/e2016-60588-6} {\bibfield
  {journal} {\bibinfo  {journal} {Eur. Phys. J. B}\ }\textbf {\bibinfo {volume}
  {89}},\ \bibinfo {pages} {41} (\bibinfo {year} {2016})}\BibitemShut {NoStop}%
\bibitem [{\citenamefont {{Wa{\ss}er}}\ \emph {et~al.}(2016)\citenamefont
  {{Wa{\ss}er}}, \citenamefont {{Lee}}, \citenamefont {{Kihou}}, \citenamefont
  {{Steffens}}, \citenamefont {{Schmalzl}}, \citenamefont {{Qureshi}},\ and\
  \citenamefont {{Braden}}}]{Waser2016}%
  \BibitemOpen
  \bibfield  {author} {\bibinfo {author} {\bibfnamefont {F.}~\bibnamefont
  {{Wa{\ss}er}}}, \bibinfo {author} {\bibfnamefont {C.~H.}\ \bibnamefont
  {{Lee}}}, \bibinfo {author} {\bibfnamefont {K.}~\bibnamefont {{Kihou}}},
  \bibinfo {author} {\bibfnamefont {P.}~\bibnamefont {{Steffens}}}, \bibinfo
  {author} {\bibfnamefont {K.}~\bibnamefont {{Schmalzl}}}, \bibinfo {author}
  {\bibfnamefont {N.}~\bibnamefont {{Qureshi}}}, \ and\ \bibinfo {author}
  {\bibfnamefont {M.}~\bibnamefont {{Braden}}},\ }\href@noop {} {\bibfield
  {journal} {\bibinfo  {journal} {ArXiv e-prints}\ } (\bibinfo {year}
  {2016})},\ \Eprint {http://arxiv.org/abs/1609.02027} {arXiv:1609.02027
  [cond-mat.supr-con]} \BibitemShut {NoStop}%
\bibitem [{not()}]{note}%
  \BibitemOpen
  \href@noop {} {}\bibinfo {note} {Jain et al. report evidence for a
  longitudinal mode appearing at 52~meV at the FM zone center. $\textbf
  Q=(2,2,0)$ is the smallest scattering vector for which an energy transfer of
  52~meV at a FM zone-center is not prohibited by kinematics in our setup using
  $k_f=4.1$\AA$^{-1}$. A smaller \textbf Q is however required due to the
  magnetic form factor in order to detect weak magnetic signals. In spite of
  very long counting we could not pick up a magnetic signal at this position.
  We also tried to find an almost equivalent longitudinal magnetic signal at
  $\textbf Q=(2.2,0,0)$, where the longitudinal branch proposed in ref.
  \cite{Jain} should not posses an energy far from 52~meV and exhibit about
  half of the strength of the transversal branch at 45~meV at $\textbf
  Q=(0,2,0)$, which we found to have 150(40)~Counts/(mon~500000). At $\textbf
  Q=(2.2,0,0)$ we could not pick up a sizable magnetic signal [13(34) and
  16(39)~Counts/(mon.~500000) for M$_y$ and M$_z$, respectively]. So we cannot
  deliver experimental hints for this longitudinal mode, but we cannot exclude
  its presence due to the big errorbars and the different \textbf Q position.
  Note that one data point in one spin flip channel is measured for 50
  minutes.}\BibitemShut {Stop}%
\bibitem [{\citenamefont {Blaha}\ \emph {et~al.}(2001)\citenamefont {Blaha},
  \citenamefont {Schwarz}, \citenamefont {Madsen}, \citenamefont {Kvasnicka},\
  and\ \citenamefont {Luitz}}]{Blaha2001}%
  \BibitemOpen
  \bibfield  {author} {\bibinfo {author} {\bibfnamefont {P.}~\bibnamefont
  {Blaha}}, \bibinfo {author} {\bibfnamefont {K.}~\bibnamefont {Schwarz}},
  \bibinfo {author} {\bibfnamefont {G.}~\bibnamefont {Madsen}}, \bibinfo
  {author} {\bibfnamefont {D.}~\bibnamefont {Kvasnicka}}, \ and\ \bibinfo
  {author} {\bibfnamefont {J.}~\bibnamefont {Luitz}},\ }\href@noop {} {\emph
  {\bibinfo {title} {WIEN2k, An Augmented Plane Wave + Local Orbitals Program
  for Calculating Crystal Properties}}}\ (\bibinfo  {publisher} {Techn.
  Universit{\"{a}}t Wien},\ \bibinfo {address} {Wien},\ \bibinfo {year}
  {2001})\BibitemShut {NoStop}%
\bibitem [{\citenamefont {Perdew}\ \emph {et~al.}(1996)\citenamefont {Perdew},
  \citenamefont {Burke},\ and\ \citenamefont {Ernzerhof}}]{Perdew1996}%
  \BibitemOpen
  \bibfield  {author} {\bibinfo {author} {\bibfnamefont {J.~P.}\ \bibnamefont
  {Perdew}}, \bibinfo {author} {\bibfnamefont {K.}~\bibnamefont {Burke}}, \
  and\ \bibinfo {author} {\bibfnamefont {M.}~\bibnamefont {Ernzerhof}},\ }\href
  {http://www.ncbi.nlm.nih.gov/pubmed/10062328} {\bibfield  {journal} {\bibinfo
   {journal} {Phys. Rev. Lett.}\ }\textbf {\bibinfo {volume} {77}},\ \bibinfo
  {pages} {3865} (\bibinfo {year} {1996})}\BibitemShut {NoStop}%
\bibitem [{\citenamefont {Lee}\ \emph {et~al.}(2006)\citenamefont {Lee},
  \citenamefont {Park}, \citenamefont {Adroja}, \citenamefont {Khomskii},
  \citenamefont {Streltsov}, \citenamefont {McEwen}, \citenamefont {Sakai},
  \citenamefont {Yoshimura}, \citenamefont {Anisimov}, \citenamefont {Mori},
  \citenamefont {Kanno},\ and\ \citenamefont {Ibberson}}]{Lee2006a}%
  \BibitemOpen
  \bibfield  {author} {\bibinfo {author} {\bibfnamefont {S.}~\bibnamefont
  {Lee}}, \bibinfo {author} {\bibfnamefont {J.-G.}\ \bibnamefont {Park}},
  \bibinfo {author} {\bibfnamefont {D.~T.}\ \bibnamefont {Adroja}}, \bibinfo
  {author} {\bibfnamefont {D.}~\bibnamefont {Khomskii}}, \bibinfo {author}
  {\bibfnamefont {S.}~\bibnamefont {Streltsov}}, \bibinfo {author}
  {\bibfnamefont {K.~A.}\ \bibnamefont {McEwen}}, \bibinfo {author}
  {\bibfnamefont {H.}~\bibnamefont {Sakai}}, \bibinfo {author} {\bibfnamefont
  {K.}~\bibnamefont {Yoshimura}}, \bibinfo {author} {\bibfnamefont {V.~I.}\
  \bibnamefont {Anisimov}}, \bibinfo {author} {\bibfnamefont {D.}~\bibnamefont
  {Mori}}, \bibinfo {author} {\bibfnamefont {R.}~\bibnamefont {Kanno}}, \ and\
  \bibinfo {author} {\bibfnamefont {R.}~\bibnamefont {Ibberson}},\ }\href
  {\doibase 10.1038/nmat1605} {\bibfield  {journal} {\bibinfo  {journal}
  {Nature materials}\ }\textbf {\bibinfo {volume} {5}},\ \bibinfo {pages} {471}
  (\bibinfo {year} {2006})}\BibitemShut {NoStop}%
\bibitem [{\citenamefont {Streltsov}\ and\ \citenamefont
  {Khomskii}(2012)}]{Streltsov2012a}%
  \BibitemOpen
  \bibfield  {author} {\bibinfo {author} {\bibfnamefont {S.~V.}\ \bibnamefont
  {Streltsov}}\ and\ \bibinfo {author} {\bibfnamefont {D.~I.}\ \bibnamefont
  {Khomskii}},\ }\href {\doibase 10.1103/PhysRevB.86.064429} {\bibfield
  {journal} {\bibinfo  {journal} {Physical Review B}\ }\textbf {\bibinfo
  {volume} {86}},\ \bibinfo {pages} {064429} (\bibinfo {year}
  {2012})}\BibitemShut {NoStop}%
\bibitem [{\citenamefont {Anisimov}\ \emph {et~al.}(1997)\citenamefont
  {Anisimov}, \citenamefont {Aryasetiawan},\ and\ \citenamefont
  {Lichtenstein}}]{Anisimov1997}%
  \BibitemOpen
  \bibfield  {author} {\bibinfo {author} {\bibfnamefont {V.~I.}\ \bibnamefont
  {Anisimov}}, \bibinfo {author} {\bibfnamefont {F.}~\bibnamefont
  {Aryasetiawan}}, \ and\ \bibinfo {author} {\bibfnamefont {A.~I.}\
  \bibnamefont {Lichtenstein}},\ }\href
  {http://iopscience.iop.org/0953-8984/9/4/002} {\bibfield  {journal} {\bibinfo
   {journal} {J. Phys.: Condens. Matter}\ }\textbf {\bibinfo {volume} {9}},\
  \bibinfo {pages} {767} (\bibinfo {year} {1997})}\BibitemShut {NoStop}%
\bibitem{pavarini} G. Zhang and E. Pavarini, Phys. Rev. B {\bf 95}, 075145 (2017).
\bibitem [{\citenamefont {Mazin}\ and\ \citenamefont
  {Singh}(1997)}]{Mazin1997}%
  \BibitemOpen
  \bibfield  {author} {\bibinfo {author} {\bibfnamefont {I.~I.}\ \bibnamefont
  {Mazin}}\ and\ \bibinfo {author} {\bibfnamefont {D.~J.}\ \bibnamefont
  {Singh}},\ }\href {\doibase 10.1103/PhysRevB.56.2556} {\bibfield  {journal}
  {\bibinfo  {journal} {Phys. Rev. B}\ }\textbf {\bibinfo {volume} {56}},\
  \bibinfo {pages} {2556} (\bibinfo {year} {1997})}\BibitemShut {NoStop}%
\bibitem [{\citenamefont {Gukasov}\ \emph {et~al.}(2002)\citenamefont
  {Gukasov}, \citenamefont {Braden}, \citenamefont {Papoular}, \citenamefont
  {Nakatsuji},\ and\ \citenamefont {Maeno}}]{Gukasov2002}%
  \BibitemOpen
  \bibfield  {author} {\bibinfo {author} {\bibfnamefont {A.}~\bibnamefont
  {Gukasov}}, \bibinfo {author} {\bibfnamefont {M.}~\bibnamefont {Braden}},
  \bibinfo {author} {\bibfnamefont {R.~J.}\ \bibnamefont {Papoular}}, \bibinfo
  {author} {\bibfnamefont {S.}~\bibnamefont {Nakatsuji}}, \ and\ \bibinfo
  {author} {\bibfnamefont {Y.}~\bibnamefont {Maeno}},\ }\href {\doibase
  10.1103/PhysRevLett.89.087202} {\bibfield  {journal} {\bibinfo  {journal}
  {Phys. Rev. Lett.}\ }\textbf {\bibinfo {volume} {89}},\ \bibinfo {pages}
  {087202} (\bibinfo {year} {2002})}\BibitemShut {NoStop}%
\bibitem [{\citenamefont {Harrison}(1999)}]{Harrison1999}%
  \BibitemOpen
  \bibfield  {author} {\bibinfo {author} {\bibfnamefont {W.~A.}\ \bibnamefont
  {Harrison}},\ }\href@noop {} {\emph {\bibinfo {title} {Elementary Electronic
  Structure}}}\ (\bibinfo  {publisher} {World Scientific},\ \bibinfo {address}
  {Singapore},\ \bibinfo {year} {1999})\ p.\ \bibinfo {pages} {817}\BibitemShut
  {NoStop}%
\bibitem [{\citenamefont {Pet\v{r}\'{i}\v{c}ek}\ \emph
  {et~al.}(2004)\citenamefont {Pet\v{r}\'{i}\v{c}ek}, \citenamefont
  {Du\v{s}ek},\ and\ \citenamefont {Palatinus}}]{Petricek}%
  \BibitemOpen
  \bibfield  {author} {\bibinfo {author} {\bibfnamefont {V.}~\bibnamefont
  {Pet\v{r}\'{i}\v{c}ek}}, \bibinfo {author} {\bibfnamefont {M.}~\bibnamefont
  {Du\v{s}ek}}, \ and\ \bibinfo {author} {\bibfnamefont {L.}~\bibnamefont
  {Palatinus}},\ }\href@noop {} {\bibfield  {journal} {\bibinfo  {journal}
  {Zeitschrift f\"ur Kristallographie - Crystalline Materials.}\ }\textbf
  {\bibinfo {volume} {229}},\ \bibinfo {pages} {345} (\bibinfo {year}
  {2004})}\BibitemShut {NoStop}%
\end{thebibliography}
\end{document}